\documentclass{aa}  

\usepackage{graphicx}
\usepackage{lscape}
\usepackage{longtable}
\usepackage{txfonts}
\begin{document}

\title{Origin of central abundances in the hot intra-cluster medium}
\subtitle{II. Chemical enrichment and supernova yield models}

\author{F. Mernier\inst{\ref{SRON},\ref{Leiden}} \and J. de Plaa\inst{\ref{SRON}} \and C. Pinto\inst{\ref{Cambridge}}  \and J. S. Kaastra\inst{\ref{SRON},\ref{Leiden}} \and P. Kosec\inst{\ref{Cambridge}} \and Y.-Y. Zhang\inst{\ref{Bonn}} \and J. Mao\inst{\ref{SRON},\ref{Leiden}} \and N. Werner\inst{\ref{Stanford1},\ref{Stanford2}} \and O. R. Pols\inst{\ref{Nijmegen}} \and J. Vink\inst{\ref{UvA}}}

\institute{SRON Netherlands Institute for Space Research, Sorbonnelaan 2, 3584 CA Utrecht, The Netherlands \\ \email{F.Mernier@sron.nl}\label{SRON} \and Leiden Observatory, Leiden University, P.O. Box 9513, 2300 RA Leiden,The Netherlands\label{Leiden} \and Institute of Astronomy, Madingley Road, CB3 0HA Cambridge, United Kingdom\label{Cambridge} \and Argelander-Institut f\"{u}r Astronomie, Auf dem H\"{u}gel 71, D-53121 Bonn, Germany\label{Bonn} \and Kavli Institute for Particle Astrophysics and Cosmology, Stanford University, 452 Lomita Mall, Stanford, CA 94305, USA\label{Stanford1} \and Department of Physics, Stanford University, 382 Via Pueblo Mall, Stanford,CA 94305-4060, USA\label{Stanford2} \and Department of Astrophysics/IMAPP, Radboud University Nijmegen, PO Box 9010, 6500 GL Nijmegen, The Netherlands\label{Nijmegen} \and Anton Pannekoek Institute for Astronomy, University of Amsterdam, Science Park 904, 1098 XH Amsterdam, The Netherlands\label{UvA}}

\date{Received 21 April 2016 / Accepted 13 July 2016}

\abstract{The hot intra-cluster medium (ICM) is rich in metals, which are synthesised by supernovae (SNe) and accumulate over time into the deep gravitational potential well of clusters of galaxies. Since most of the elements visible in X-rays are formed by type Ia (SNIa) and/or core-collapse (SNcc) supernovae, measuring their abundances gives us direct information on the nucleosynthesis products of billions of SNe since the epoch of the star formation peak ($z \sim $ 2--3). In this study, we compare the most accurate average X/Fe abundance ratios (compiled in a previous work from \textit{XMM-Newton} EPIC and RGS observations of 44 galaxy clusters, groups, and ellipticals), representative of the chemical enrichment in the nearby ICM, to various SNIa and SNcc nucleosynthesis models found in the literature. The use of a SNcc model combined to any favoured standard SNIa model (deflagration or delayed-detonation)  fails to reproduce our abundance pattern. In particular, the Ca/Fe and Ni/Fe ratios are significantly underestimated by the models. We show that the Ca/Fe ratio can be reproduced better, either by taking a SNIa delayed-detonation model that matches the observations of the Tycho supernova remnant, or by adding a contribution from the  ``Ca-rich gap transient'' SNe, whose material should easily mix into the hot ICM. On the other hand, the Ni/Fe ratio can be reproduced better by assuming that both deflagration and delayed-detonation SNIa contribute in similar proportions to the ICM enrichment. In either case, the fraction of SNIa over the total number of SNe (SNIa+SNcc) contributing to the ICM enrichment ranges within 29--45\%. This fraction is found to be systematically higher than the corresponding SNIa/(SNIa+SNcc) fraction contributing to the enrichment of the proto-solar environnement (15--25\%). We also  discuss and quantify two useful constraints on both SNIa (i.e. the initial metallicity on SNIa progenitors and the fraction of low-mass stars that result in SNIa) and SNcc (i.e. the effect of the IMF and the possible contribution of pair-instability SNe to the enrichment) that can be inferred from the ICM abundance ratios. Finally, we show that detonative sub-Chandrasekhar WD explosions (resulting, for example, from violent WD mergers) cannot be a dominant channel for SNIa progenitors in galaxy clusters.}


\maketitle

\titlerunning{Origin of central abundances in the hot intra-cluster medium II}
\authorrunning{F. Mernier et al.}

\section{Introduction}\label{sect:intro}

Since the emergence and progress of stellar nucleosynthesis models over the past century, it is now well known that all the heavy elements in the Universe (i.e. except H, He, and traces of Li and Be, which were  produced shortly after the Big Bang)  have been produced by stars and stellar remnants \citep{1957AJ.....62....9C,1957RvMP...29..547B}. In particular, $\alpha$- and Fe-peak elements ($8 \le \text{Z} \le 28$) are mostly synthesised by nuclear fusion reactions during stellar lifetimes and supernova (SN) explosions, and are then released into and beyond the interstellar medium \citep[e.g.][]{1973ARA&A..11...73A,1980FCPh....5..287T}. On the one hand, oxygen (O), neon (Ne), magnesium (Mg), silicon (Si), and sulfur (S), are thought to be mostly produced by core-collapse supernovae (SNcc). On the other hand, Type Ia supernovae (SNIa) produce predominantly argon (Ar), calcium (Ca), chromium (Cr), manganese (Mn), iron (Fe), and nickel (Ni). Finally, when low-mass stars (<6 $M_\sun$) leave the main sequence and enter into their asymptotic giant branch (AGB) phase, they are efficient in releasing lighter metals, such as carbon (C) or nitrogen (N), via powerful winds. Although this general picture of synthesis (and recycling) of metals through cosmic ages is now well established, many issues are still unsolved and still bring  a great deal of  uncertainty when identifying the specific origins of each element.

First, it is well known that SNcc result from the end-of-life explosion of massive stars ($\sim$10--140 $M_\sun$). However, several parameters, such as the mass cut that separates the collapsing core from the supernova remnant (SNR) or the final kinetic energy of the explosion, are poorly constrained. Consequently, some differences in the predicted abundance pattern from proposed SNcc models proposed by different groups still remain \cite[for a review, see][]{2013ARA&A..51..457N}. Moreover, the relative amount of unburned elements depends on the initial mass and metallicity of the progenitor star \citep[e.g.][]{1995ApJS..101..181W}. These two parameters are not always easy to constrain, especially when considering a whole population of (massive) stars, whereas the universality of the initial mass function (IMF) is still under debate \citep[e.g.][]{2010ApJ...709.1195T,2012MNRAS.422L..33D}.

Second, despite their fundamental role both in the Galactic chemical evolution \citep[e.g.][]{1995ApJS...98..617T,2009ApJ...707.1466K} and as standard candles for cosmological distances \citep{1998AJ....116.1009R,1998ApJ...507...46S}, the nature of SNIa progenitors is still elusive \citep[for reviews, see][]{2011NatCo...2E.350H,2012PASA...29..447M,2013FrPhy...8..116H,2014ARA&A..52..107M}. It seems likely that the explosion results from an accreting carbon-oxygen white dwarf (WD), which ignites shortly before reaching its Chandrasekhar mass. However, it is not clear whether the mass transfer is due to a normal stellar companion \citep[single degenerate scenario;][]{1973ApJ...186.1007W} or a second white dwarf \citep[double degenerate scenario;][]{1984ApJ...277..355W,1984ApJS...54..335I}. Furthermore, the physics of the SNIa explosion itself is poorly constrained \citep[for a review, see][]{2000ARA&A..38..191H}. In most models, the explosion starts with a deflagration (i.e. the burning front propagates subsonically). The currently favoured explosion models suggest that when the burning front reaches a certain critical density, it propagates supersonically, and the deflagration becomes a detonation. These so-called delayed-detonation models \citep{1989MNRAS.239..785K}, in particular their variant of deflagration-to-detonation transition (DDT), have been studied in detail \citep[e.g.]{1991A&A...245L..25K,1997ApJ...475..740N,2005ApJ...623..337G,2013MNRAS.429.1156S}, but not yet fully understood. Moreover, some peculiar SNIa \citep[e.g. the 2002cx supernovae,][]{2013MNRAS.429.2287K} seem to be better explained by invoking a pure deflagration explosion \citep{2004ApJ...606..413B,2006AJ....132..189J,2007PASP..119..360P}. What is clear, however, is that the abundance pattern of the elements synthesised by SNIa is very sensitive to their explosion mechanism.

Many attempts to constrain all these SNcc and SNIa uncertainties have been made by studying the optical and X-ray spectra of SNRs and, particularly, their abundance pattern \citep[e.g.][]{2006ApJ...645.1373B,2014PASJ...66...68Y}. However, such an approach is difficult in practice, mostly because  only a few Galactic SNRs are suitable for studying the composition of their ejecta, preventing any statistical study over large samples;  the emitting plasma of the SNRs is far from being in ionisation equilibrium, making its spectroscopy complicated and not yet fully understood; and  the ejected material from the SNR easily mixes with the surrounding ISM, making it a challenge to correctly estimate  the metal abundances from the SN itself. \\

An interesting alternative approach to investigating nucleosynthesis products from supernovae (SNe) is to consider the chemical enrichment at the scale of galaxy clusters. In fact, the hot intra-cluster medium (ICM) pervading the volume of galaxy clusters and groups, and accounting for no less than $\sim$80\% of their total baryonic matter, is rich in $\alpha$- and Fe peak elements \citep[for reviews, see][]{2008SSRv..134..337W,2010A&ARv..18..127B}. These metals, which can be observed via their emission lines from X-ray spectroscopy, must have been synthesised by SNIa and SNcc inside the cluster galaxies, and have enriched the ICM, especially around $z\simeq$ 2--3, during the major cosmic epoch of star formation \citep{2006ApJ...651..142H,2014ARA&A..52..415M}. Assuming that the large gravitational potential well of clusters/groups make them behave like a closed-box system, the metal abundances of the ICM are a remarkable signature of the yields of billions of SNIa and SNcc over time. Moreover, the ICM is well known to be in (or very close to) collisional ionisation equilibrium state, making its spectroscopy less complex than SN spectra and its abundances relatively easy to derive.

Several previous studies have already attempted to use abundance measurements in the ICM in order to constrain SNIa and SNcc yield models. 
For instance, \citet{2007A&A...465..345D} compiled a sample of 22 cool-core clusters, and found that the standard SNIa models fail to reproduce the Ar/Ca and Ca/Fe abundance ratios. They also showed that the fraction of SNIa over the total number of SNe highly depends of the considered models. \citet{2009A&A...508..565D} showed that Si/Fe abundance ratios are remarkably uniform over a sample of 26 cool-core clusters observed with \textit{XMM-Newton}, arguing for a similar enrichment process within cluster cores. However, they concluded that systematic uncertainties between the SN models are too large to precisely estimate the relative contribution of SNIa and SNcc. Finally, many abundance studies have been performed on individual objects as well \citep[e.g.][]{2006A&A...449..475W,2006A&A...452..397D,2007ApJ...667L..41S,2009A&A...493..409S,2015A&A...575A..37M}. From these studies, and considering the instrumental performance of current X-ray observatories, it appears that higher quality data (i.e. with longer exposure time), collected over larger samples, are needed to clarify the picture of the precise origin of metals in the ICM.

In this paper, we make use of ICM abundances measured in two previous works \citep[de Plaa et al., to be submitted;][hereafter Paper I]{2016arXiv160601165M} and compare them with predictions from theoretical SNIa and SNcc yield models. These measurements consist of the average X/Fe abundance ratios of ten elements (O/Fe, Ne/Fe, Mg/Fe, Si/Fe, S/Fe, Ar/Fe, Ca/Fe, Cr/Fe, Mn/Fe, and Ni/Fe) in the ICM of 44 cool-core clusters, groups, and ellipticals, using the \textit{XMM-Newton} EPIC and RGS instruments. To our knowledge, this is the most complete and robust abundance pattern measured in the ICM  available to date. 

This paper is structured as follows.
In Sect. \ref{sect:data_reduction} we present the sample and briefly recall the data reduction, as well as the spectral analysis used to derive the abundance ratios. We then discuss the comparison between various SNIa and SNcc models and our average ICM abundance pattern (Sect. \ref{sect:ICM}) on the one hand, and  the proto-solar abundances (Sect. \ref{sect:proto-solar}) on the other hand. Section \ref{sect:conclusion} summarises our discussion and addresses future prospects.
Throughout this paper we assume cosmological parameters of $H_0 = 70$ km s$^{-1}$ Mpc$^{-1}$, $\Omega_m = 0.3,$ and $\Omega_\Lambda = 0.7$. The abundances presented in this paper are taken relative to the proto-solar values of \citet{2009LanB...4B...44L}. All the error bars are given at a $68\%$ confidence level.


\section{Observations and spectral analysis}\label{sect:data_reduction}

\begin{figure}[!]
        \centering
                \includegraphics[width=0.49\textwidth]{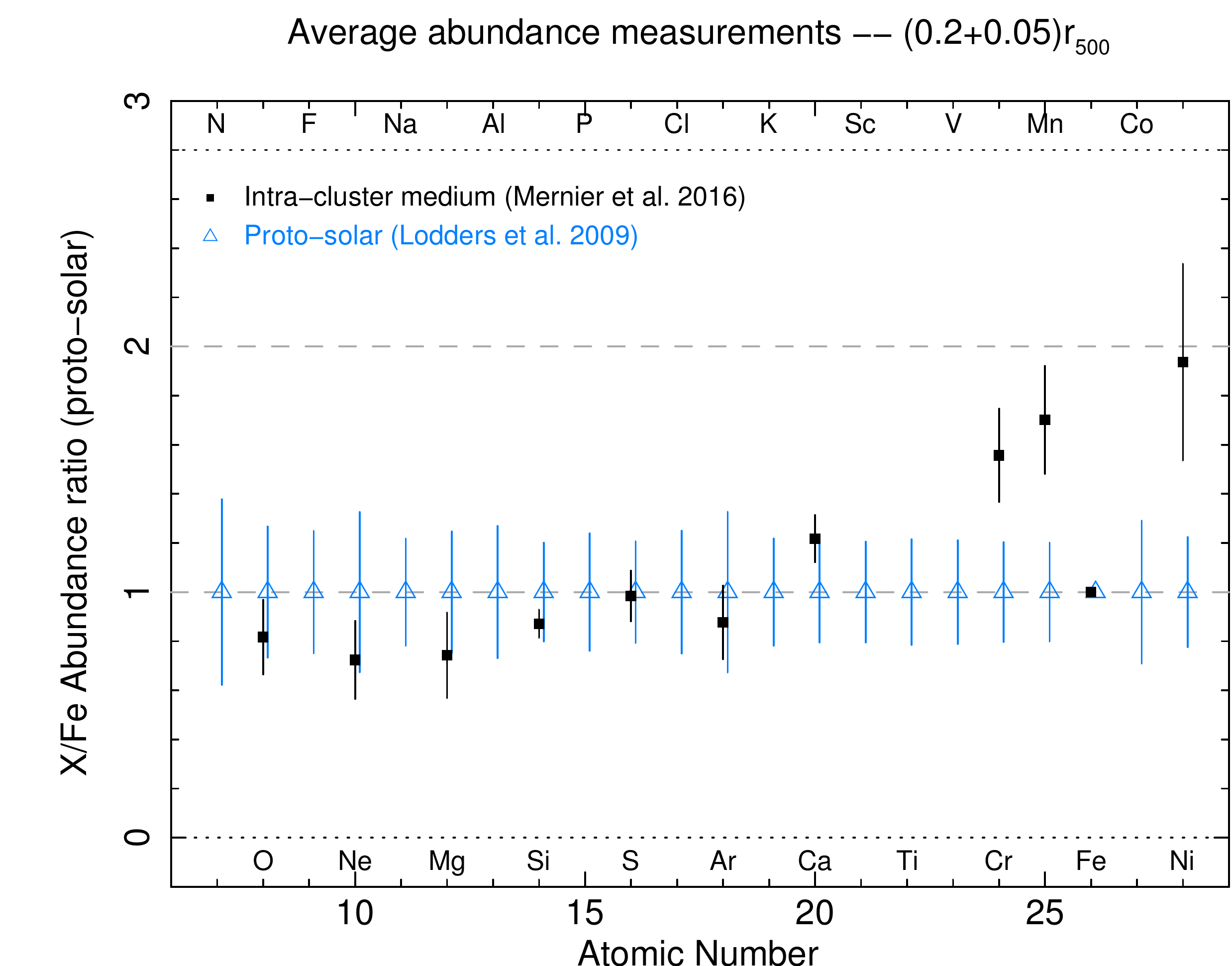}

        \caption{Average abundance ratios measured in the ICM (black filled squares, see Paper I) versus proto-solar abundances \citep[blue empty triangles, adapted from][]{2009LanB...4B...44L} and their 1$\sigma$ uncertainties. }
\label{fig:cluster_vs_solar}
\end{figure}

We start by briefly summarising the main steps of the data reduction and the spectral analysis that were necessary to provide the average X/Fe abundance pattern (Fig. \ref{fig:cluster_vs_solar}), representative of the ICM of cool-core objects. The detailed presentation, reduction, and spectral analysis of our observations can be found in Paper I and in de Plaa et al. (to be submitted).

Our sample consists of the CHEERS\footnote{CHEmical Enrichment Rgs Sample} catalogue (de Plaa et al., to be submitted), and is detailed in Table 1 in Paper I \citep[see also][de Plaa et al., to be submitted]{2015A&A...575A..38P}. It includes 44 nearby ($z < 0.1$) cool-core clusters, groups, and elliptical galaxies for which the \ion{O}{viii} 1s--2p line at 19 $\AA$  is detected by the RGS instrument with >5$\sigma$. Recent \textit{XMM-Newton} observations (AO-12, PI: de Plaa) have been combined with archival data. We reduced the EPIC and RGS data using the \textit{XMM-Newton} Science Analysis System (SAS)  software v14.0.0. After having filtered them from solar-flare events, we obtain cleaned EPIC MOS\,1, MOS\,2, and pn data of $\sim$4.5, $\sim$4.6, and $\sim$3.7 Ms, respectively.

The EPIC spectra were extracted within a circle of a radius of either $0.2 r_{500}$ (for $kT_\text{mean}$ > 1.7 keV, i.e. the farther clusters) or $0.05 R_{500}$ (for $kT_\text{mean}$ < 1.7 keV, i.e. the nearer groups/ellipticals). The RGS spectra were extracted with a cross-dispersion width of 0.8$'$. We carefully checked that the difference in these EPIC and RGS extraction regions did not affect our final results (Paper I).

We used the SPEX fitting package \citep{1996uxsa.conf..411K} v2.05 to perform our spectral fits. The EPIC and RGS spectra were fitted with a \texttt{gdem} and a 2T (i.e. \texttt{cie+cie}) thermal model, respectively. The EPIC (X-ray and particle) background components were carefully modelled following the method detailed in \citet{2015A&A...575A..37M}. The free parameters in our fits were the emission measure (or normalisation), the temperature parameters ($kT_\text{mean}$ and $\sigma_T$ for a \texttt{gdem} model; $kT_\text{up}$ and $kT_\text{low}$ for a 2T model), and the abundances of O, Ne, Mg, Si, S, Ar, Ca, Fe, and Ni for EPIC, and O, Ne, Mg, and Fe for RGS. We were also able to constrain the EPIC sample-averaged abundances of Cr and Mn by converting the equivalent width of their line fluxes (at rest-frame energies of $\sim$5.7 keV and $\sim$6.2 keV, respectively) as described in \citet{2015A&A...575A..37M}.

We finally computed a weighted average of all the considered X/Fe abundance ratios (taking O/Fe and Ne/Fe from RGS, and Mg/Fe, Si/Fe, S/Fe, Ar/Fe, Ca/Fe, Cr/Fe, Mn/Fe, and Ni/Fe from EPIC), carefully taking account of all the possible systematic uncertainties that could affect our measurements. This final abundance pattern, reasonably representative of the ICM enrichment in the cool cores of clusters, groups, and ellipticals, is shown in Fig. \ref{fig:cluster_vs_solar} (see also Fig. 8 in Paper I). The Cr/Fe, Mn/Fe, and Ni/Fe abundance ratios are found to differ significantly from the proto-solar values. The ICM average abundance pattern, as well as the proto-solar estimates\footnote{The proto-solar abundances used in this paper \citep{2009LanB...4B...44L} are  currently the most representative  abundances of the solar system at its formation as they are based on meteoritic compositions.} and their uncertainties \citep{2009LanB...4B...44L}, can now be directly compared to various sets of SN yield models.

\section{Chemical enrichment in the ICM}\label{sect:ICM}


Since the metals present in the ICM are the product of billions of SNe that exploded mostly in the cluster galaxies, the average abundance ratios measured in the ICM  bear witness to the contribution of both SNIa and SNcc to the chemical enrichment of galaxy clusters and groups. Following several past attempts \citep{2006A&A...449..475W,2006A&A...452..397D,2007A&A...465..345D,2015A&A...575A..37M}, we fit a combination of SNIa and SNcc nucleosynthesis models to our ICM average abundance pattern. More quantitatively, the total number of atoms of the $i$-th element in the ICM can be expressed as a linear combination of the number of atoms expected from SNIa ($N_{i,\text{Ia}}$) and SNcc ($N_{i,\text{cc}}$) contributions \citep[e.g.][]{2006A&A...449..475W}
\begin{equation}\label{eq:lin_com_SNe}
N_{i,\text{tot}} = a \, N_{i,\text{Ia}} + b \, N_{i,\text{cc}},
\end{equation}
where $a$ and $b$ are multiplicative factors corresponding respectively to the number of SNIa and SNcc that released their metal contents into the ICM. Since $N_{i,\text{Ia}}$ and $N_{i,\text{cc}}$ can be easily converted into abundances, we can fit this linear combination to our average abundance pattern (ten data points), and infer the SNIa-to-SNe fraction
\begin{equation}\label{eq:SNe_ratio}
\frac{\text{SNIa}}{\text{SNIa} + \text{SNcc}},
\end{equation}
which represents the relative number of SNIa over the total number of SNe responsible for the enrichment. As noted by \citet{2005PASA...22...49M}, the equations above assume an instantaneous recycling of the metals, and such a ratio should not be interpreted as the true relative number of SNIa over the entire lifetime of the clusters, but rather as the SNIa ratio necessary to enrich the ICM \citep{2007A&A...465..345D}.

\begin{figure*}[!]
        \centering
                \includegraphics[width=0.49\textwidth]{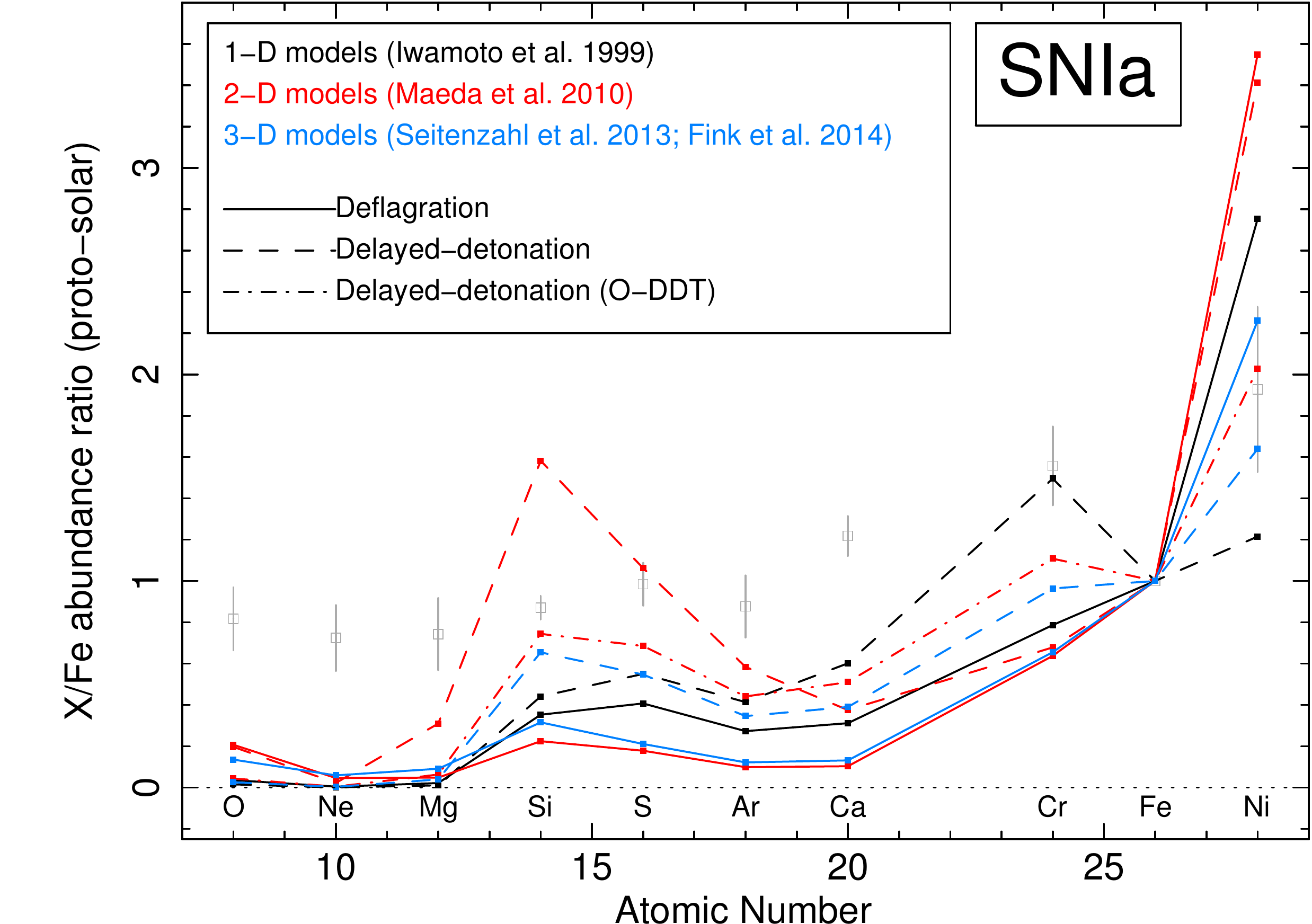}
                \includegraphics[width=0.49\textwidth]{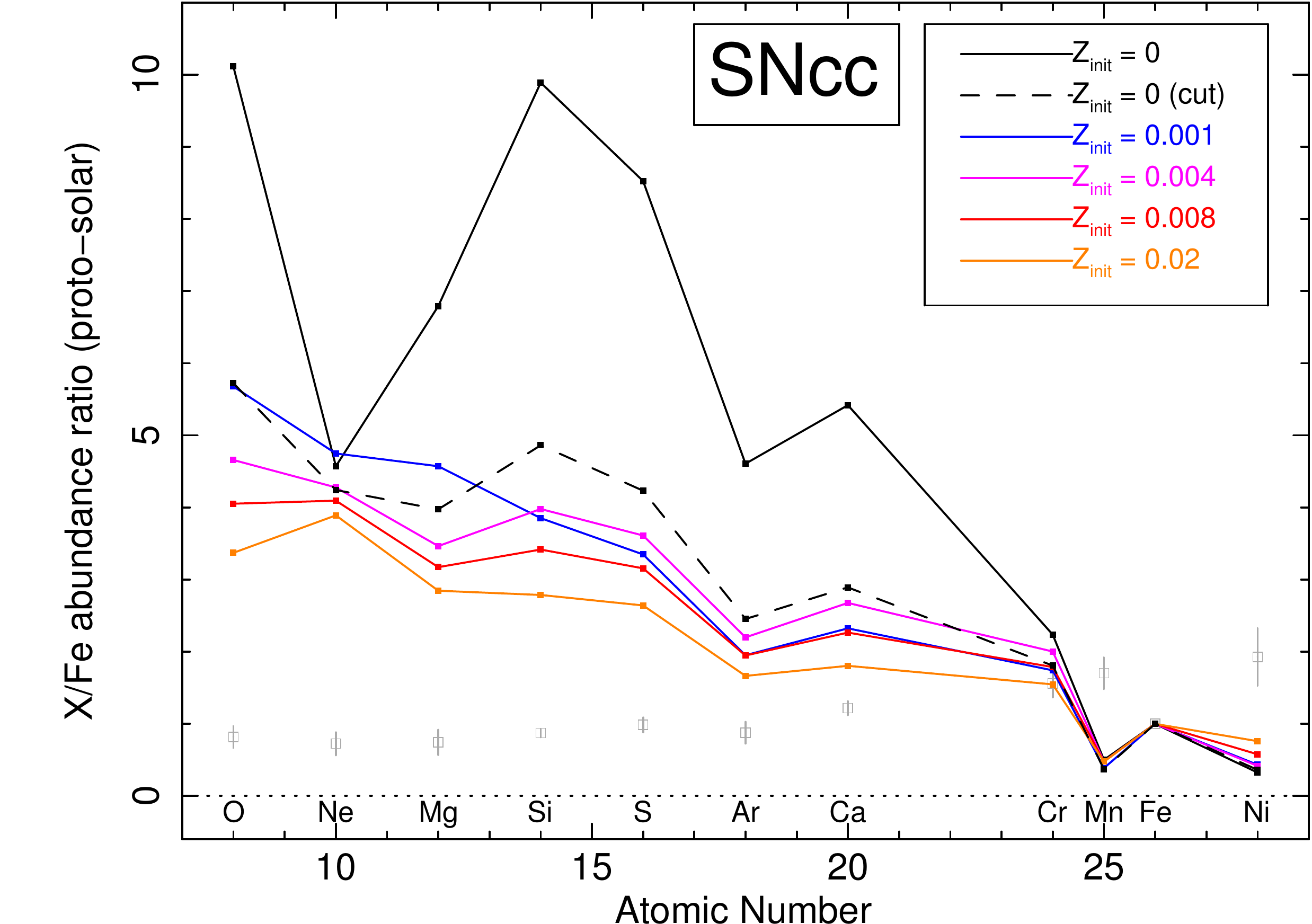}

        \caption{Predicted X/Fe abundances from various SNIa and SNcc yield models. For comparison, the ICM average abundance ratios (inferred from Paper I) are also plotted. \textit{Left:} SNIa yield models: 1-D \citep[W7 and WDD2 from][]{1999ApJS..125..439I}, 2-D \citep[C-DEF, C-DDT, and O-DDT from][]{2010ApJ...712..624M}, and 3-D (N100def and N100 from \citet{2014MNRAS.438.1762F} and \citet{2013MNRAS.429.1156S}, respectively) models are indicated in back, red, and blue, respectively. A distinction is also made between the  explosion models: deflagration (W7, C-DEF, and N100def; solid lines) and  delayed-detonation (dashed lines for WDD2, C-DDT, and N100; dash-dotted lines for O-DDT). The Mn/Fe ratio is not shown here because it highly depends on the initial metallicity of SNIa progenitors (Sect. \ref{sect:manganese}). \textit{Right:} SNcc models, all taken from \citet{2013ARA&A..51..457N}, assuming a Salpeter IMF. Various initial metallicities ($Z_\text{init}$) for SNcc progenitors are compared. The dashed line corresponds to the Z0\_cut model (see text).}
\label{fig:SNe_models}
\end{figure*}

Throughout this paper, many SNIa and SNcc yield models are considered. They are all summarised in Table \ref{table:SNe_models}, and described further in the text when needed. In Fig. \ref{fig:SNe_models}, we plot the X/Fe abundance pattern predicted from several individual SNIa (left panel) and SNcc (right panel) models. In particular, we emphasise the differences in the nucleosynthesis of SNIa deflagration and delayed-detonation explosions (left panel), and the effects of the initial metallicity ($Z_\text{init}$) of massive stars on their predicted SNcc yields (right panel). Specific comparisons are also discussed  in this paper.

\subsection{Abundance pattern of even-Z elements}\label{sect:even}

In this section, we consider only the ratio of even-Z elements (i.e. O/Fe, Ne/Fe, Mg/Fe, Si/Fe, S/Fe, Ar/Fe, Ca/Fe, Cr/Fe, and Ni/Fe) as part of the ICM abundance pattern. In fact, the Mn/Fe ratio is particular, in the sense that it may depend on the metallicity of the SNIa progenitors, which has not been fully taken into account in most of the yield models so far. For this reason, Mn/Fe needs to be considered separately. In Sect. \ref{sect:manganese}, we discuss this initial metallicity dependence extensively and  we derive other useful information related to SNIa progenitors in general from the
observed Mn/Fe ratio.

\subsubsection{Classical SNIa and Nomoto SNcc yields}\label{sect:SNe_models}

One set of SNIa models commonly referred to in the literature (hereafter the ``Classical'' models, Table \ref{table:SNe_models}) is from \citet{1999ApJS..125..439I}, who predicted nucleosynthesis products regarding different one-dimensional (1-D) explosion mechanisms. Two initial central densities ($\rho_{9}$, given in units of $10^9$ g/cm$^3$) are considered (C and W models, see Table \ref{table:SNe_models}). The W7 and W70 models assume a pure deflagration during the SNIa event, while the WDD and CDD models assume a delayed-detonation, with three possible transition densities ($\rho_{T,7}$, given in units of $10^7$ g/cm$^3$). The models currently favoured  by the supernova community are the delayed-detonation models, and among these, WDD2 is usually preferred \citep{1999ApJS..125..439I}.

A well-referenced set of SNcc models was given by \citet{2006NuPhA.777..424N} and has been recently updated by \citet[][hereafter the ``Nomoto'' models]{2013ARA&A..51..457N}, who estimated nucleosynthesis products of a SNcc as a function of the mass and the initial metallicity ($Z_\text{init} = 0, 0.001, 0.004, 0.008$, or $0.02$) of its progenitor. In order to estimate the total yield mass $M_{i,\text{SNcc}}$ of the $i$-th element coming from SNcc explosions, we integrate these models \citep[following][]{1995MNRAS.277..945T} over a power-law IMF between 10--40 $M_\sun$ \citep[or 10--140 $M_\sun$ when $Z_\text{init} = 0$;][]{2013ARA&A..51..457N}, as 
\begin{equation}\label{eq:IMF_integration}
M_{i,\text{SNcc}} = \frac{\int_{10 M_\sun}^{(1)40 M_\sun} M_i (m) \ m^{-(1+x)} \ dm}{\int_{10 M_\sun}^{(1)40 M_\sun} m^{-(1+x)} \ dm},
\end{equation}
where $M_i (m)$ is the mass yield of the $i$-th element at a given mass $m$ of the main sequence progenitor and $x$ is the power index of the IMF. Here we assume that the fraction of metals resulting from the SNcc enrichment have been generated by a population of massive stars having a Salpeter IMF ($x = 1.35$) and sharing a common $Z_\text{init}$. We note that in the case $Z_\text{init} = 0$, the stellar yields beyond 40 $M_\sun$ are available for 100 $M_\sun$ and 140 $M_\sun$ only. Consequently, a precise integration over the IMF (Eq. \ref{eq:IMF_integration}) within the 40--140 $M_\sun$ range is not trivial, and the IMF-weighted abundance ratios of the Z0 model might be somewhat altered by the choice of the mass binning. For this reason, in the following we also consider the Z0\_cut model, similar to the Z0 model, but restricted to $\le$40 $M_\sun$ \citep[Table \ref{table:SNe_models} and Fig. \ref{fig:SNe_models} left; see also][]{2006NuPhA.777..424N}.

\begin{figure}[!]
        \centering
                \includegraphics[width=0.49\textwidth]{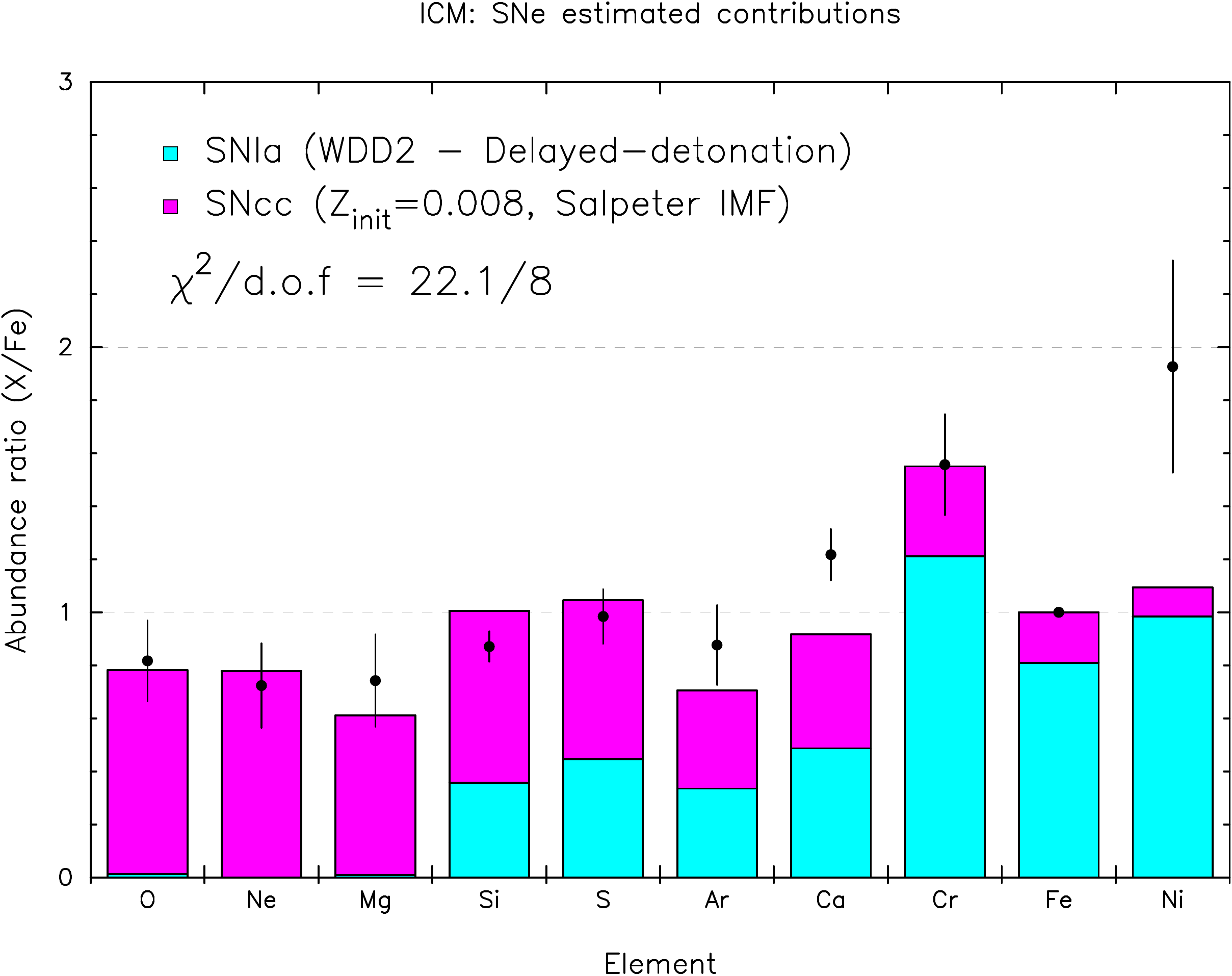}

        \caption{Average abundance ratios versus atomic numbers in the average ICM abundance pattern (Paper I). The histograms show the yields contribution of a best-fit combination of one Classical SNIa model (WDD2) and one Nomoto SNcc ($Z_\text{init} = 0.008$, and Salpeter IMF) model.}
\label{fig:SNe_fits_classical}
\end{figure}

Considering all the possible combinations of a Classical SNIa model plus a Nomoto SNcc model, our best fit (Fig. \ref{fig:SNe_fits_classical}) is reached for a WDD2 model and a SNcc initial metallicity $Z_\text{init}=0.008$. With a reduced $\chi^2$ of $\sim$2.8, this fit is quite poor. In particular, the Ar/Fe, Ca/Fe, and Ni/Fe ratios are underestimated with >1$\sigma$, >3$\sigma$, and >2$\sigma$ respectively, while Si/Fe is overestimated with >2$\sigma$. In Table \ref{table:SNe_fits} (top panel) we indicate the five best fits (including this one) that we find with this Classical combination of models, as well as their respective estimated SN fractions. The errors on the SN fractions are typically about $\pm$5--6\%. Although all these fits are poor, the delayed-detonation models are always favoured. The SNIa rates are always comparable, ranging between $\sim$29\% and $\sim$35\%. We note that the WDD2 model with $Z=0.02$ used in \citet{2007A&A...465..345D} has a reduced $\chi^2$ of $\sim$2.9, and also shows clear discrepancies in Ca/Fe (>3$\sigma$), and Ni/Fe (>2$\sigma$). Clearly, these combinations fail to reproduce our measured ICM abundance pattern.

\subsubsection{ Ca/Fe ratio: A contribution from Ca-rich SNe?}\label{sect:calcium}

In previous studies and in this work the Ca/Fe measured ratio in the ICM was found to be higher than expected. Using the same  Classical models \citep[together with the SNcc models of][]{2006NuPhA.777..424N}, \citet{2006A&A...449..475W}, \citet{2006A&A...452..397D}, \citet{2007A&A...465..345D}, and \citet{2015A&A...575A..37M} reported a significant underestimate of the Ca/Fe expected yields compared to the measurements. Also in stellar populations, chemical evolution models from \citet{2013arXiv1312.0606C} (who made use of the Classical SNIa yields predictions as well) failed to reproduce observations of Ca/Fe. Although residual systematic biases in the Ca abundance measurements cannot be excluded, this possibility is quite unlikely. Indeed, atomic databases have been considerably improved during the past decades, the continuum and line fluxes in this work have been fitted carefully, the EPIC MOS and pn instruments agree very well in their Ca/Fe measurements, and no line feature around $\sim$4 keV has been reported so far in the EPIC effective areas or in the particle background (for more details, see Paper I). 

\begin{figure*}[!]
        \centering
                \includegraphics[width=0.49\textwidth]{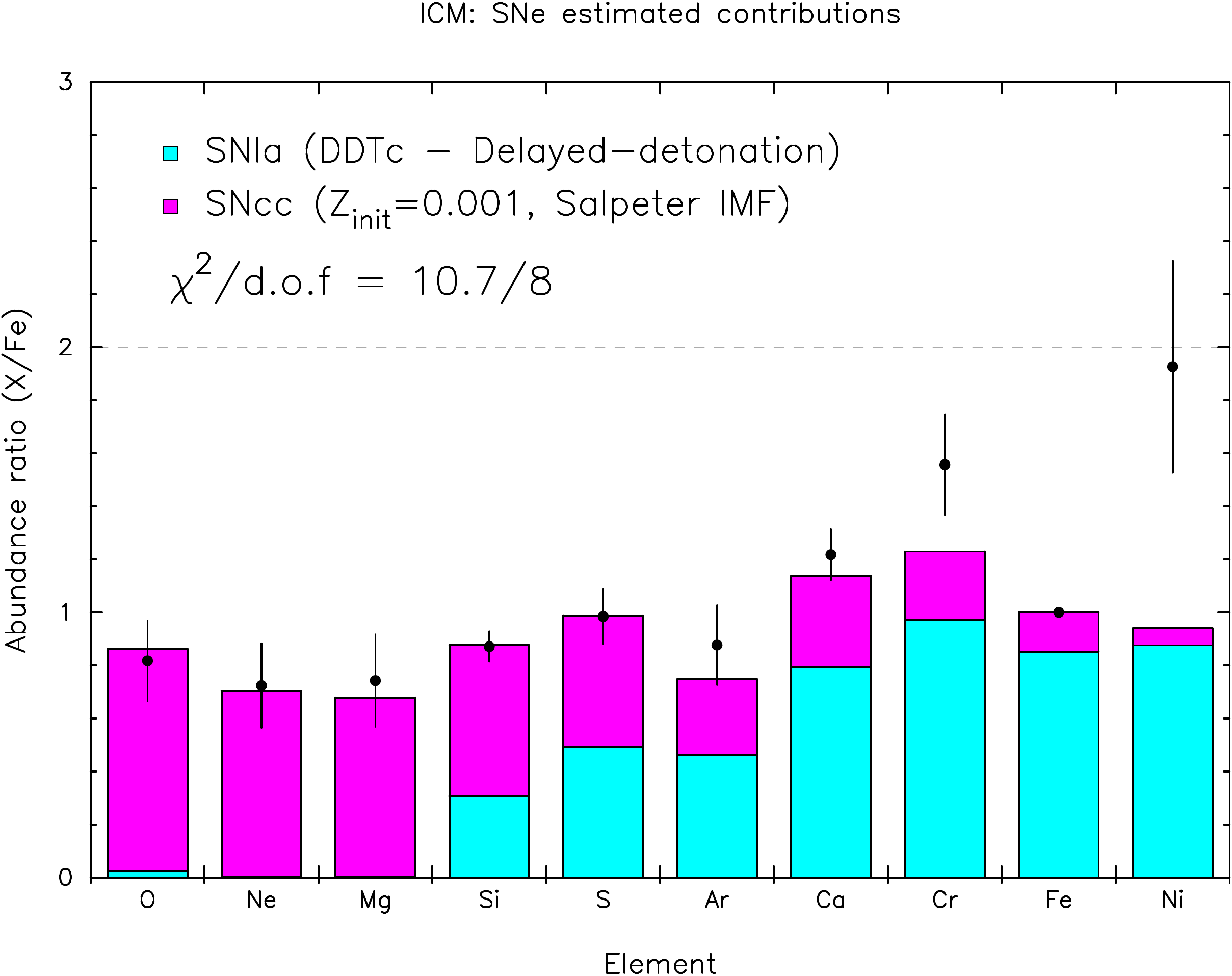}
                \includegraphics[width=0.49\textwidth]{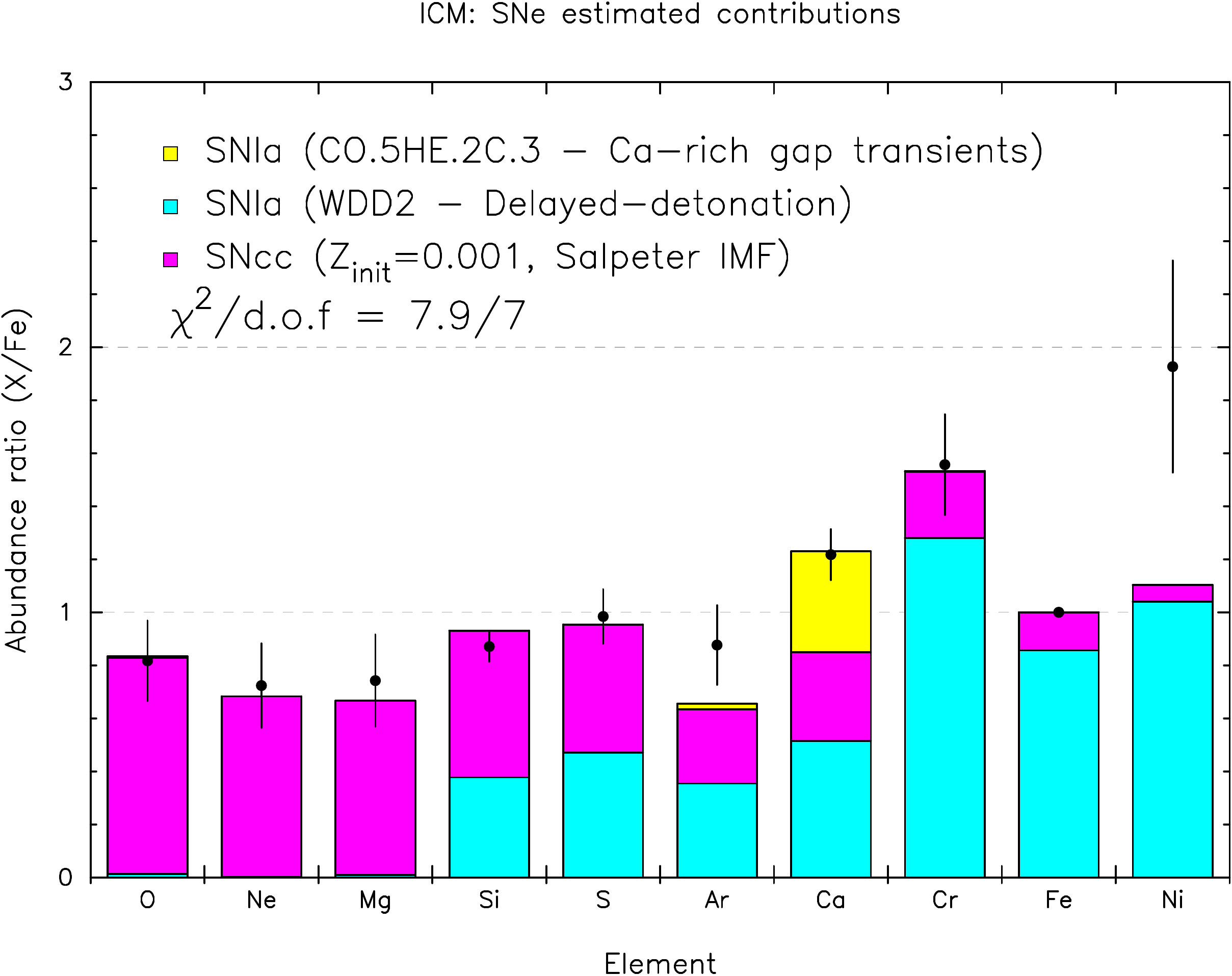} \\
                \includegraphics[width=0.49\textwidth]{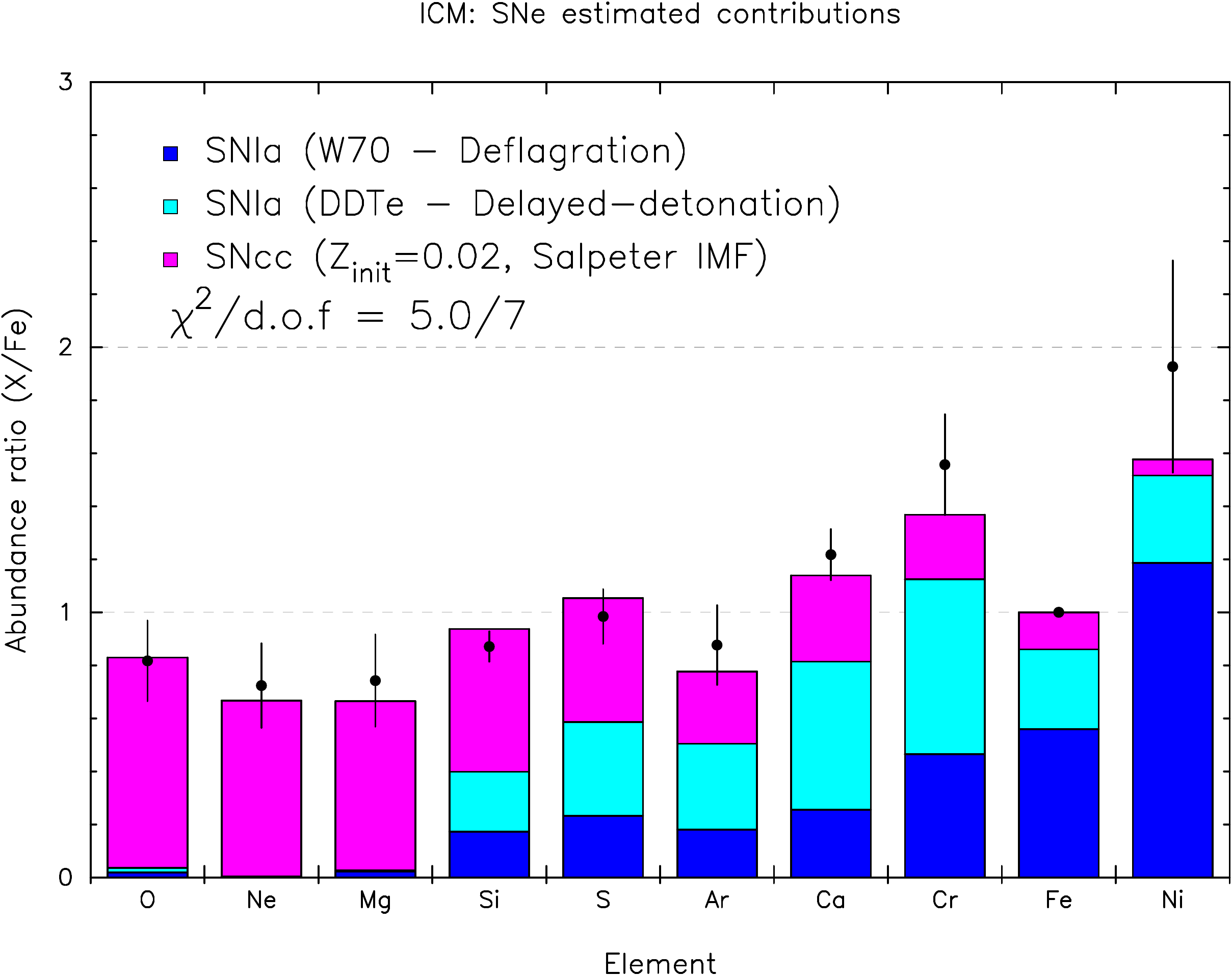}
                \includegraphics[width=0.49\textwidth]{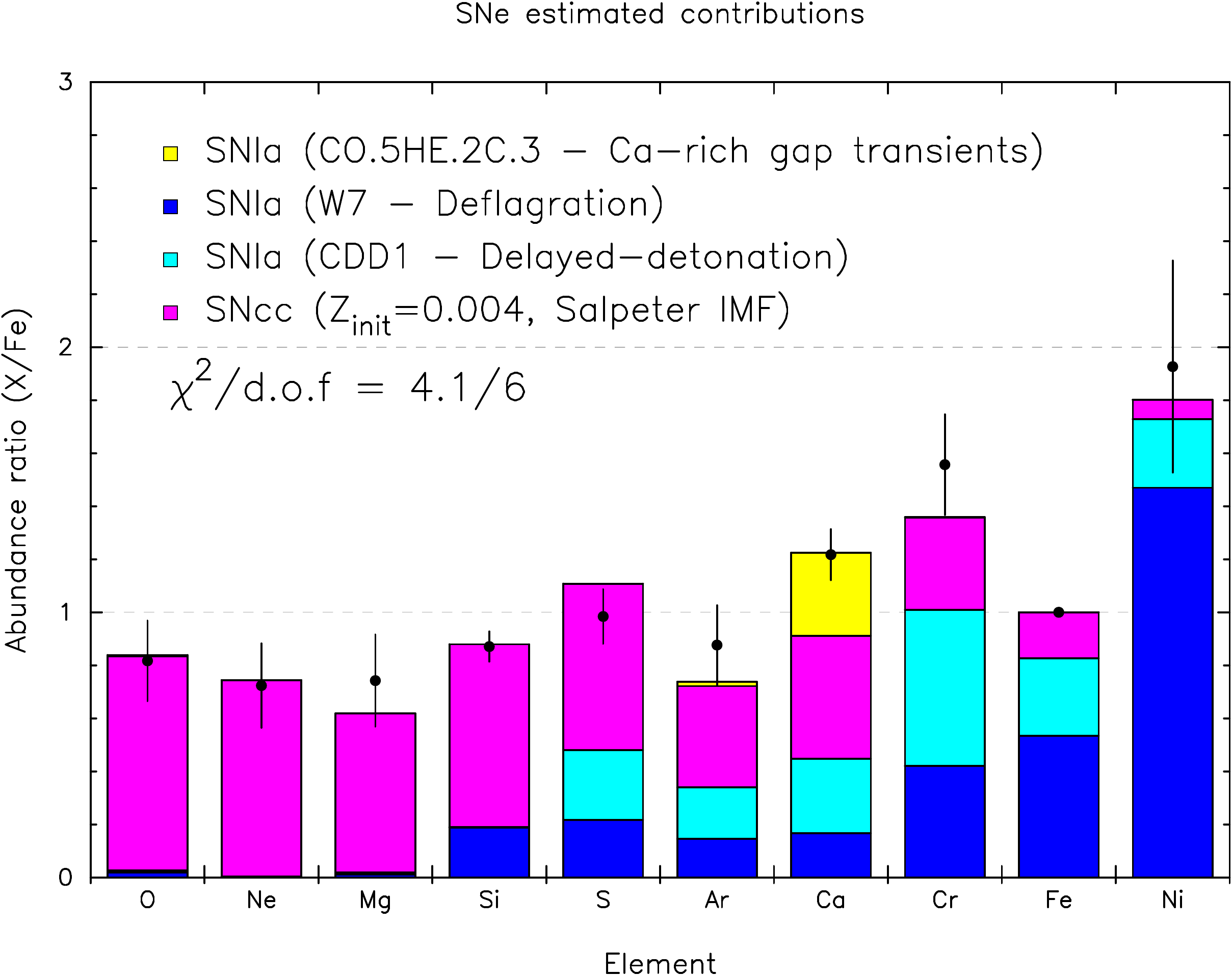}

        \caption{Same as Fig. \ref{fig:SNe_fits_classical}, but fitting alternative sets of models. In every case, only one SNcc model (Nomoto)  has been fitted ($Z_\textit{init} = 0, 0.001, 0.004, 0.008$ or 0.02; Salpeter IMF). \textit{Top left}: Delayed-detonation SNIa model based on the observation of Tycho supernova remnant (Bravo, DDTc). \textit{Top right}: combination of a Classical delayed-detonation SNIa model (WDD2) and a Ca-rich gap transients population model (CO.5HE.2C.3). \textit{Bottom left}: combination of a Classical deflagration SNIa model (W70) and a delayed-detonation SNIa model based on the observation of Tycho SN remnant (Bravo, DDTe). \textit{Bottom right}: combination of a Classical deflagration SNIa model (W7), a Ca-rich gap transients population model (CO.5HE.2C.3), and a Classical delayed-detonation SNIa model (CDD1). }
\label{fig:SNe_fits_others}
\end{figure*}

As one possibility of solving this conundrum, \citet{2007A&A...465..345D} made use of an alternative delayed-detonation SNIa model, which provides the best description of the spectra of the Tycho SNR \citep{2006ApJ...645.1373B}. More specifically, \citet{2007A&A...465..345D} showed that the DDTc model of \citet{1996A&A...306..811B} better fits the measured Ar/Fe and Ca/Fe ratios. In this paper, we test three models (DDTa, DDTc, and DDTe; hereafter the "Bravo" models) introduced in \citet{2003ApJ...593..358B} and \citet{2006ApJ...645.1373B} that are based on the calculations of \citet[][]{1996A&A...306..811B} and that reasonably reproduce the spectral features of Tycho. The best fit using these models is shown in Fig. \ref{fig:SNe_fits_others} (top left) and Table \ref{table:SNe_fits} (second panel).

Similarly to \citet{2007A&A...465..345D}, we obtain the best fit to our abundance pattern when using the DDTc model. The fits are significantly better than in the Classical models (for the best fit, $\chi^2/\text{d.o.f.} \simeq 1.3$), essentially because this alternative successfully reproduces the observed Si/Fe, Ar/Fe, and (above all) Ca/Fe ratios. However, the Ni/Fe ratio is still clearly underestimated (>2$\sigma$). The SNIa-to-SNe fraction ranges from $\sim$29\% to $\sim$35\%, which is similar to what was found for the Nomoto+Classical case.

Another possibility has been recently proposed by \citet{2014ApJ...780L..34M}, and suggests a significant additional contribution from Ca-rich gap transients to the ICM enrichment. Although spectroscopically defined as Type Ib/Ic, this recently discovered subclass of SNe \citep{2003IAUC.8159....2F,2010Natur.465..322P,2011ApJ...728L..36P} is thought to originate from a He-accreting WD \citep{2011ApJ...738...21W,2015MNRAS.452.2463F} rather than a core-collapse object, and will be further considered as being part of the SNIa contribution. Their nebular spectrum is dominated by Ca, they show large photospheric velocities \citep{2012ApJ...755..161K}, and they preferentially explode far from galaxies \citep{2013MNRAS.432.1680Y}, likely making their nucleosynthesis products easily mixed into the ICM (see below). We explore this possibility by adding one "Ca-rich gap" yield model to the Classical SNIa and Nomoto SNcc models. We base this additional contribution on a set of yield models calculated by \citet{2011ApJ...738...21W}, who considered various masses of the CO core and the He upper layer of the accreting WD (as well as, for instance, a 2\% mass fraction of N in the He layer, or a mixing of 30\% between the CO core and the He layer). The decimal numbers in the model acronyms refer to the mass of each considered layer \citep[in $M_\sun$; see][and Table \ref{table:SNe_models}]{2011ApJ...738...21W}.

The best fit is obtained for a Z0.008+W70+CO.5HE.2N.02 combination, with a reduced $\chi^2$ of $\sim$0.7. Compared to the Nomoto+Classical models, the fit is thus significantly improved and fully acceptable. The enriching fraction of SNIa over the total number of SNe is estimated to be $\sim$40\%, thus similar to (although somewhat higher than) what was found in the previous cases.

However, based on this best fit, we estimate that the relative fraction of Ca-rich gap transients over the total number of SNIa contributing to the enrichment, $\text{SNIa(Ca)}/\text{SNIa}$, is $\sim$34\%. This is much larger than recent estimates of the Ca-rich SNe rate over the total SNIa rate from the literature; i.e. $7 \pm 5 \%$ \citep{2010Natur.465..322P}, <20\% \citep{2011MNRAS.412.1473L}, and $\sim$16\% \citep{2014ApJ...780L..34M}. Since Ca-rich gap transients occur preferably in the outskirts of galaxies (or even in the intra-cluster light) and have large photospheric velocities (see above), one interesting possibility is
that  they may be significantly more efficient in enriching the ICM than classical SNIa (whose metals may be more easily locked in the gravitational well of galaxy members). The fraction $\text{SNIa(Ca)}/\text{SNIa}$ contributing to the enrichment might thus naturally be higher than the absolute Ca-rich/SNIa rate (for comparison with the solar neighbourhood enrichment, see also Sect. \ref{sect:proto-solar}). 
On the other hand, the amount of produced Ca highly depends on the models. In particular, assuming 30\% of mixing between the CO core and the He layer in the accreting WD produces significantly more Ca during the explosion \citep{2011ApJ...738...21W}, and thus requires a smaller contribution from Ca-rich gap transients to the total enrichment. Considering this particular case (i.e. taking the CO.5HE.2C.3 model only; see Table \ref{table:SNe_fits}, third panel), the best fit is achieved for the combination Z0.001+WDD2+CO.5HE.2C.3 (with a reduced $\chi^2$ of $\sim$1.1, thus formally acceptable as well), which is plotted in Fig. \ref{fig:SNe_fits_others} (top right). We then obtain $\text{SNIa(Ca)}/\text{SNIa} \simeq 9$\%, which is in agreement with the estimated rates from the literature. The enriching SNIa-to-SNe fraction is $\sim$35--40\%. It is important to note, in this case, the need for a SNIa delayed-detonation model (WDD2) instead of a deflagration model, in order to predict a consistent Cr/Fe ratio\footnote{In addition to Ca, the CO.5HE.2N.02 model produces a significant fraction of Cr. The fits then favour SNIa deflagration models because in compensation they predict a limited Cr/Fe ratio and match the high observed Ni/Fe ratio. On the contrary, the CO.5HE.2C.3 model does not produce Cr, and the Cr/Fe ratio can only be successfully reproduced by using a delayed-detonation model for the SNIa contribution.}. However, the use of a delayed-detonation model again underestimates the Ni/Fe ratio, as noted previously (Fig. \ref{fig:SNe_fits_classical} and \ref{fig:SNe_fits_others}, top left). This shows that the choice of a specific Ca-rich gap contribution has a significant impact on favouring one of the two possible SNIa explosion mechanisms. A further discussion of the Ni/Fe ratio and the choice of a SNIa explosion model can be found in Sect. \ref{sect:nickel}. 

At this stage, we must point out that the assumption of a significant fraction of the ICM enrichment coming from Ca-rich gap transient SNe is purely speculative. However, as demonstrated in this section, this may be a realistic possibility, as it successfully reproduces the high Ca/Fe abundance ratio measured in the ICM. Whereas in the rest of our analysis we choose to constrain the most realistic Ca-rich gap model regarding the current estimates of the Ca-rich/SNIa rate (i.e. allowing the CO.5HE.2C.3 model only), we must note that such rates are not well constrained yet by the observations. Alternatively, more precise measurements of the abundances in the ICM using the next generation of X-ray satellites could potentially help to constrain this rate as well (see also Sect. \ref{sect:future}).

\begin{table*}[!]
\begin{centering}
\caption{Results of various combinations of SN fits to the average ICM abundance pattern (Paper I). In each case, only one SNcc model has been fitted ($Z_\text{init} = 0, 0.001, 0.004, 0.008,$ or 0.02; Salpeter IMF), and we only show the five best fits, sorted by increasing $\chi^2 / $d.o.f. (degrees of freedom).}             
\label{table:SNe_fits}
\setlength{\tabcolsep}{11pt}
\begin{tabular}{c c c c c c c c}        
\hline \hline                

SNcc & \multicolumn{3}{c}{SNIa} & $\frac{\text{SNIa}}{\text{SNIa}+\text{SNcc}}$ & $\frac{\text{SNIa(Ca)}}{\text{SNIa}}$ & $\frac{\text{SNIa(def)}}{\text{SNIa}}$ & $\chi^2 / \text{d.o.f.}$ \\
 & & & & &  & \\
\hline
Nomoto & Classical & $-$ & $-$ &  &  &  &  \\    
\hline
Z0.008 & WDD2 &  &  & $0.31$ & $-$ & $-$ & $22.1/8$ \\
Z0.02 & WDD2 &  &  & $0.29$ & $-$ & $-$ & $22.8/8$ \\
Z0.001 & WDD2 &  &  & $0.35$ & $-$ & $-$ & $22.8/8$ \\
Z0.008 & CDD2 &  &  & $0.30$ & $-$ & $-$ & $23.0/8$ \\
Z0.004 & WDD2 &  &  & $0.29$ & $-$ & $-$ & $23.0/8$ \\
\hline
Nomoto & Bravo & $-$ & $-$ &  &  &  &  \\    
\hline
Z0.001 & DDTc &  &  & $0.35$ & $-$ & $-$ & $10.7/8$ \\
Z0.008 & DDTc &  &  & $0.32$ & $-$ & $-$ & $11.3/8$ \\
Z0.02 & DDTc &  &  & $0.29$ & $-$ & $-$ & $11.6/8$ \\
Z0.004 & DDTc &  &  & $0.33$ & $-$ & $-$ & $12.4/8$ \\
Z0\_cut & DDTc &  &  & $0.32$ & $-$ & $-$ & $16.5/8$ \\
\hline
Nomoto & Classical & Ca-rich gap & $-$ &  &  &  &  \\    
\hline
Z0.001 & WDD2 & CO.5HE.2C.3\tablefootmark{(*)} &  & $0.38$ & $0.09$ & $-$ & $7.9/7$ \\
Z0.02 & WDD2 & CO.5HE.2C.3\tablefootmark{(*)} &  & $0.33$ & $0.10$ & $-$ & $8.1/7$ \\
Z0.02 & CDD2 & CO.5HE.2C.3\tablefootmark{(*)} &  & $0.31$ & $0.11$ & $-$ & $8.3/7$ \\
Z0.001 & CDD2 & CO.5HE.2C.3\tablefootmark{(*)} &  & $0.37$ & $0.11$ & $-$ & $8.4/7$ \\
Z0.008 & WDD3 & CO.5HE.2C.3\tablefootmark{(*)} &  & $0.31$ & $0.12$ & $-$ & $9.5/7$ \\
\hline
Nomoto & Classical & Bravo & $-$ &  &  &  &  \\    
\hline
Z0.001 & W70 & DDTe &  & $0.40$ & $-$ & $0.58$ & $5.0/7$ \\
Z0.02 & W70 & DDTe &  & $0.35$ & $-$ & $0.58$ & $6.6/7$ \\
Z0.001 & W7 & DDTe &  & $0.42$ & $-$ & $0.50$ & $6.8/7$ \\
Z0.001 & WDD3 & DDTe&  & $0.38$ & $-$ & $-$ & $7.5/7$ \\
Z0.02 & W7 & DDTe &  & $0.36$ & $-$ & $0.51$ & $8.1/7$ \\
\hline
Nomoto & Classical & Classical & Ca-rich gap &  &  &  &  \\    
\hline
Z0.004 & W7 & CDD1 & CO.5HE.2C.3\tablefootmark{(*)} & $0.36$ & $0.07$ & $0.57$ & $4.1/6$ \\
Z0.008 & W7 & CDD1 & CO.5HE.2C.3\tablefootmark{(*)} & $0.35$ & $0.08$ & $0.56$ & $4.1/6$ \\
Z0.004 & W70 & CDD1 & CO.5HE.2C.3\tablefootmark{(*)} & $0.35$ & $0.07$ & $0.70$ & $5.5/6$ \\
Z0.008 & W70 & CDD1 & CO.5HE.2C.3\tablefootmark{(*)} & $0.34$ & $0.08$ & $0.68$ & $5.6/6$ \\
Z0\_cut & W7 & CDD1 & CO.5HE.2C.3\tablefootmark{(*)} & $0.35$ & $0.07$ & $0.54$ & $5.9/6$ \\

\hline                                   
\end{tabular}
\par\end{centering}
\tablefoot{The choice of the CO.5HE.2C.3 model, indicated by a \textit{(*)}, has been fixed (see text).}
\end{table*}

\subsubsection{ Ni/Fe ratio: Diversity in SNIa explosions?}\label{sect:nickel}

During SNcc explosions most of the Ni remains locked in the collapsing core, while in SNIa explosions the Ni production depends on the electron capture efficiency in the core. In particular, delayed-detonation models (i.e. the models currently favoured  by the supernova community) should produce limited amounts of Ni. \citet{2000ApJ...528..139D} used \textit{ASCA} to measure a large Ni/Fe abundance ratio of $\sim$4 in the central region of three clusters. They deduced that this ratio is more consistent with SNIa deflagration models and inconsistent with delayed-detonation models. \citet{2005AdSpR..36..677B} measured the abundances of O, Si, and Fe in four clusters, and favour predictions from WDD models. Other papers based on the abundance ratios of more Si-group elements \citep{2007A&A...465..345D,2015A&A...575A..37M} also show a better consistency with the delayed-detonation models. However, when measured, the Ni/Fe ratio is very often found to be super-solar \citep[e.g.][Paper I]{2009ApJ...705L..62T,2009A&A...508..565D}. Comparing their Ni/Fe value of $1.5 \pm 0.3$ solar with the ones found by \citet{2000ApJ...528..139D}, \citet{2002A&A...381...21F} suggested that both deflagration and delayed-detonation SNIa could participate in the enrichment of the ICM.

We explore this compromise by modelling one additional Classical contribution of SNIa to our two combinations already described in Sect. \ref{sect:calcium}. Again, we choose to use the CO.5HE.2C.3 model as the most reasonable possibility for the Ca-rich gap contribution (Sect. \ref{sect:calcium}). The five best fits of the Nomoto+Classical+Bravo and Nomoto+Classical+Classical+Ca-rich gap models are presented in Table \ref{table:SNe_fits} (last two  panels). The best fits of these two cases are shown in Fig. \ref{fig:SNe_fits_others} (bottom left and bottom right, respectively).
These two combinations of models are now fully consistent ($\lesssim$1$\sigma$) with all our average abundance ratios. With a reduced $\chi^2$ of $\sim$0.7 in both cases, the fits are better than all our previous attempts discussed above. From Table \ref{table:SNe_fits} (last two  panels), it also appears that at least four out of the five best fits of these combinations include a contribution of one deflagration and one delayed-detonation model. The relative number of deflagration SNIa over the total number of SNIa contributing to the enrichment, $\text{SNIa(\text{def})}/\text{SNIa}$, is typically  in the range of  50--70\%. It is, however, very difficult to discriminate between the best fits of each case. Similarly, we cannot clearly favour either of the two cases above since their respective best fits reproduce the average abundance pattern equally well within the uncertainties (Fig. \ref{fig:SNe_fits_others}, lower panels). However, no matter which case we select, again the SNIa-to-SNe fraction (34--42\%) is comparable with the estimates in the cases discussed earlier.

Such a possibility for a SNIa bimodality in the enrichment processes of the ICM is interesting. In many respects, the bimodal nature of SNIa has already been clearly established. For instance, it seems that $\sim$50\% of SNIa explode promptly ($\sim$$10^8$ years after the starburst), while the other half explode much later, following an exponential decrease with a time scale of $\sim$3 Gy \citep{2006MNRAS.370..773M}. Furthermore, while a population of luminous SNIa with a slow magnitude decline is mostly found in late-type galaxies, another population of subluminous SNIa has a steeper decline, and seems to explode preferentially in old elliptical galaxies \citep{2000AJ....120.1479H}. It is likely that the bright, slowly declining SNIa correspond to the "prompt" population, while the subluminous and fast declining SNIa correspond to the "delayed" population.
Similarly, while the supernova community is still debating the nature of the SNIa progenitors (see also Sect. \ref{sect:SNIa_progenitors}), recent results suggest that both the single-degenerate and the double-degenerate scenarios might co-exist in nature \citep[e.g.][]{2001ApJ...546..734L,2014MNRAS.445.2535S,2015Natur.521..328C,2015Natur.521..332O} \citep[see however][]{2001PASP..113..169B}.
Considering all these indications of diversity in SNIa, a bimodal population of deflagration and delayed-detonation SNIa responsible for the enrichment of the ICM remains possible. Moreover, it should be noted that such a diversity in the explosion mechanisms of SNIa has already been proposed, from results of an optical study \citep{2000ApJ...543L..49H}. This might bring one more piece to the complex puzzle of SNIa and their progenitors.

Some alternative  scenarios to explain the high Ni/Fe ratio can be also considered. There is compelling evidence that some SNIa produce large fractions of Ni \citep[e.g.][]{2015ApJ...801L..31Y}. On the other hand, some SNcc may overproduce Ni as well, sometimes at a super-solar level \citep{2015ApJ...807..110J}, and it is possible that the current yield models actually underestimate the Ni production within SNcc. Finally, the Ni/Fe ratio from SNIa contribution may be sensitive to the initial metallicity of the SNIa progenitors (see further discussion in Sect. \ref{sect:SNIa_metallicity}).

Despite these intriguing possibilities, it is important to note that measuring the Ni abundance is a challenge  using the current X-ray capabilities. In the abundance pattern derived in Paper I and used for this work, substantial systematic uncertainties have been taken into account to overcome the large disagreement between MOS and pn. Moreover, the hard band (7--9 keV) in which the main Ni-K lines reside is often significantly contaminated by the instrumental background. Despite the very careful background modelling performed in Paper I, we cannot fully exclude that the background still affects our Ni/Fe measurements in both MOS and pn detectors (see also discussion in Paper I). Finally, the SN models themselves have uncertainties in their yield predictions (e.g. related to the electron capture rates adopted in SNIa models, see Appendix \ref{sect:ecapture}), and prevent us from firmly favouring one specific combination of models \citep{2009A&A...508..565D}. A better future  constraint on the Ni/Fe ratio, coupled to updated SNIa and SNcc yield models, will help us to favour one particular SNIa explosion model, and perhaps to confirm (or rule out) the co-existence of two explosion mechanisms (Sect. \ref{sect:future}).

\subsubsection{Two- and three-dimensional SNIa yield models}

Whereas all the nucleosynthesis yields considered so far are based on calculations assuming a 1-D (i.e. spherically symmetric) explosion, several authors have recently published various sets of SNIa yields, assuming two-dimensional (2-D) or even three-dimensional (3-D) explosions. In this section, we  compare these updated yields with our observations in order to determine whether predictions from multi-dimensional SNIa calculations better reproduce our ICM abundance pattern.

We take the 2-D models (deflagration and delayed-detonation)  from \citet{2010ApJ...712..624M}, as well as the 3-D delayed-detonation and the 3-D deflagration models from \citet{2013MNRAS.429.1156S} and \citet{2014MNRAS.438.1762F}, respectively. These models (hereafter  "2D" and "3D") are mentioned in Table \ref{table:SNe_models} (see also Fig. \ref{fig:SNe_models}, left panel). In addition to the symmetrical (deflagration and delayed-detonation) cases, the 2D models also propose an asymmetrical delayed-detonation explosion (O-DDT), where the ignition is slightly offset from the WD centre. In the 3D models, various numbers of ignition spots (usually close to the WD centre) are successively considered, sometimes with changing values for $\rho_9$. In order to check whether such multi-dimensional models better agree with our ICM abundance pattern, we re-fit our results, this time replacing the Classical (1-D) models successively by the 2D and the 3D models. The full results are shown in Tables \ref{table:SNe_fits_2D} and \ref{table:SNe_fits_3D} (for the 2D and 3D cases, respectively). We note that the available Bravo and Ca-rich gap models have only been calculated for one dimension so far, so we could not apply any 2-D or 3-D extensions to those.

\begin{table*}[!]
\begin{centering}
\caption{Same as Table \ref{table:SNe_fits}, but considering 2-D SNIa models instead of the 1-D Classical SNIa models.}             
\label{table:SNe_fits_2D}
\setlength{\tabcolsep}{11pt}
\begin{tabular}{c c c c c c c c}        
\hline \hline                

SNcc & \multicolumn{3}{c}{SNIa} & $\frac{\text{SNIa}}{\text{SNIa}+\text{SNcc}}$ & $\frac{\text{SNIa(Ca)}}{\text{SNIa}}$ & $\frac{\text{SNIa(def)}}{\text{SNIa}}$ & $\chi^2 / \text{d.o.f.}$ \\
 & & & & &  & \\
\hline
Nomoto & 2D & $-$ & $-$ &  &  &  &  \\    
\hline
Z0.02 & O-DDT &  &  & $0.35$ & $-$ & $-$ & $56.0/8$ \\
Z0.001 & O-DDT &  &  & $0.41$ & $-$ & $-$ & $60.0/8$ \\
Z0.008 & O-DDT &  &  & $0.37$ & $-$ & $-$ & $63.1/8$ \\
Z0.004 & O-DDT &  &  & $0.39$ & $-$ & $-$ & $69.0/8$ \\
Z0.008 & C-DEF &  &  & $0.36$ & $-$ & $-$ & $79.9/8$ \\
\hline
Nomoto & 2D & Ca-rich gap & $-$ &  &  &  &  \\    
\hline
Z0.02 & O-DDT & CO.5HE.2C.3\tablefootmark{(*)} &  & $0.39$ & $0.10$ & $-$ & $30.2/7$ \\
Z0.001 & O-DDT & CO.5HE.2C.3\tablefootmark{(*)} &  & $0.45$ & $0.09$ & $-$ & $32.7/7$ \\
Z0.02 & C-DEF & CO.5HE.2C.3\tablefootmark{(*)} &  & $0.38$ & $0.09$ & $-$ & $36.8/7$ \\
Z0.008 & O-DDT & CO.5HE.2C.3\tablefootmark{(*)} &  & $0.41$ & $0.09$ & $-$ & $39.5/7$ \\
Z0.008 & C-DEF & CO.5HE.2C.3\tablefootmark{(*)} &  & $0.40$ & $0.08$ & $-$ & $40.1/7$ \\
\hline
Nomoto & 2D & 2D & Ca-rich gap &  &  &  &  \\    
\hline
Z0.02 & C-DEF & O-DDT & CO.5HE.2C.3\tablefootmark{(*)} & $0.39$ & $0.10$ & $0.13$ & $26.3/6$ \\
Z0.001 & C-DEF & O-DDT & CO.5HE.2C.3\tablefootmark{(*)} & $0.45$ & $0.09$ & $0.07$ & $30.3/6$ \\
Z0.008 & C-DEF & O-DDT & CO.5HE.2C.3\tablefootmark{(*)} & $0.41$ & $0.09$ & $0.16$ & $33.9/6$ \\
Z0.004 & C-DEF & O-DDT & CO.5HE.2C.3\tablefootmark{(*)} & $0.43$ & $0.08$ & $0.18$ & $39.9/6$ \\
Z0.02 & C-DEF & C-DDT & CO.5HE.2C.3\tablefootmark{(*)} & $0.40$ & $0.09$ & $0.79$ & $41.9/6$ \\

\hline                                   
\end{tabular}
\par\end{centering}
\end{table*}

\begin{table*}[!]
\begin{centering}
\caption{Same as Table \ref{table:SNe_fits}, but considering 3-D SNIa models instead of the  1-D Classical SNIa models.}             
\label{table:SNe_fits_3D}
\setlength{\tabcolsep}{11pt}
\begin{tabular}{c c c c c c c c}        
\hline \hline                

SNcc & \multicolumn{3}{c}{SNIa} & $\frac{\text{SNIa}}{\text{SNIa}+\text{SNcc}}$ & $\frac{\text{SNIa(Ca)}}{\text{SNIa}}$ & $\frac{\text{SNIa(def)}}{\text{SNIa}}$ & $\chi^2 / \text{d.o.f.}$ \\
 & & & & &  & \\
\hline
Nomoto & 3D & $-$ & $-$ &  &  &  &  \\    
\hline
Z0.008 & N100H &  &  & $0.26$ & $-$ & $-$ & $47.2/8$ \\
Z0.004 & N100H &  &  & $0.28$ & $-$ & $-$ & $47.3/8$ \\
Z0.02 & N100H &  &  & $0.25$ & $-$ & $-$ & $49.8/8$ \\
Z0.001 & N100H &  &  & $0.30$ & $-$ & $-$ & $55.1/8$ \\
Z0.02 & N150 &  &  & $0.31$ & $-$ & $-$ & $55.6/8$ \\
\hline
Nomoto & 3D & Ca-rich gap & $-$ &  &  &  &  \\    
\hline
Z0.02 & N100H & CO.5HE.2C.3\tablefootmark{(*)} &  & $0.31$ & $0.17$ & $-$ & $11.8/7$ \\
Z0.008 & N100Hdef & CO.5HE.2C.3\tablefootmark{(*)} &  & $0.44$ & $0.11$ & $-$ & $12.0/7$ \\
Z0.02 & N100Hdef & CO.5HE.2C.3\tablefootmark{(*)} &  & $0.43$ & $0.12$ & $-$ & $12.8/7$ \\
Z0.004 & N100Hdef & CO.5HE.2C.3\tablefootmark{(*)} &  & $0.45$ & $0.10$ & $-$ & $12.9/7$ \\
Z0.008 & N100H & CO.5HE.2C.3\tablefootmark{(*)} &  & $0.32$ & $0.15$ & $-$ & $13.1/7$ \\
\hline
Nomoto & 3D & 3D & Ca-rich gap &  &  &  &  \\    
\hline
Z0.008 & N100Hdef & N100H & CO.5HE.2C.3\tablefootmark{(*)} & $0.42$ & $0.11$ & $0.77$ & $11.2/6$ \\
Z0.02 & N100Hdef & N150 & CO.5HE.2C.3\tablefootmark{(*)} & $0.41$ & $0.12$ & $0.69$ & $11.3/6$ \\
Z0.02 & N100Hdef & N100L & CO.5HE.2C.3\tablefootmark{(*)} & $0.43$ & $0.11$ & $0.76$ & $11.4/6$ \\
Z0.02 & N100Hdef & N100 & CO.5HE.2C.3\tablefootmark{(*)} & $0.41$ & $0.12$ & $0.75$ & $11.6/6$ \\
Z0.02 & N100Hdef & N100H & CO.5HE.2C.3\tablefootmark{(*)} & $0.40$ & $0.12$ & $0.76$ & $11.7/6$ \\

\hline                                   
\end{tabular}
\par\end{centering}
\end{table*}

\begin{figure*}[!]
        \centering
                \includegraphics[width=0.49\textwidth]{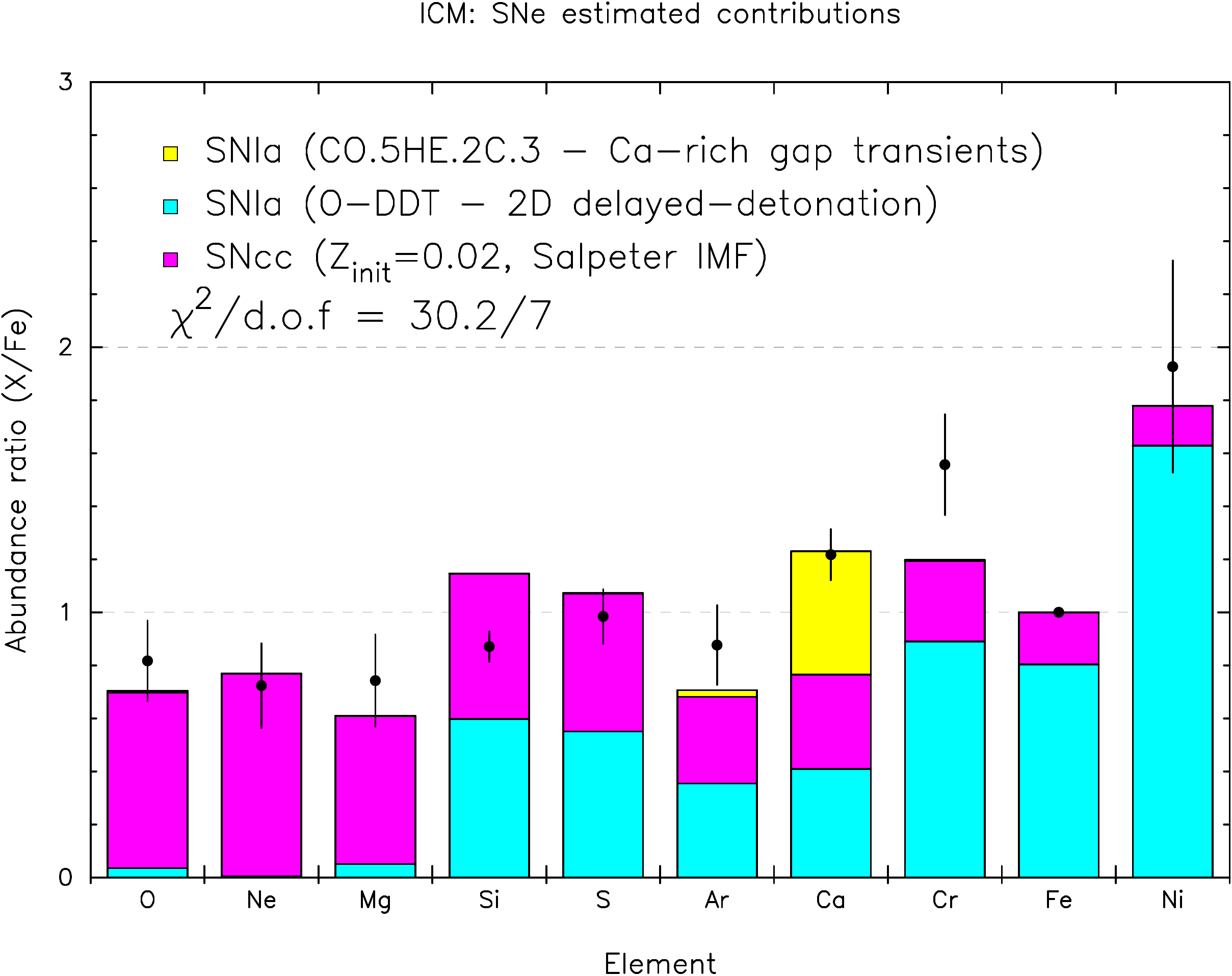}
                \includegraphics[width=0.49\textwidth]{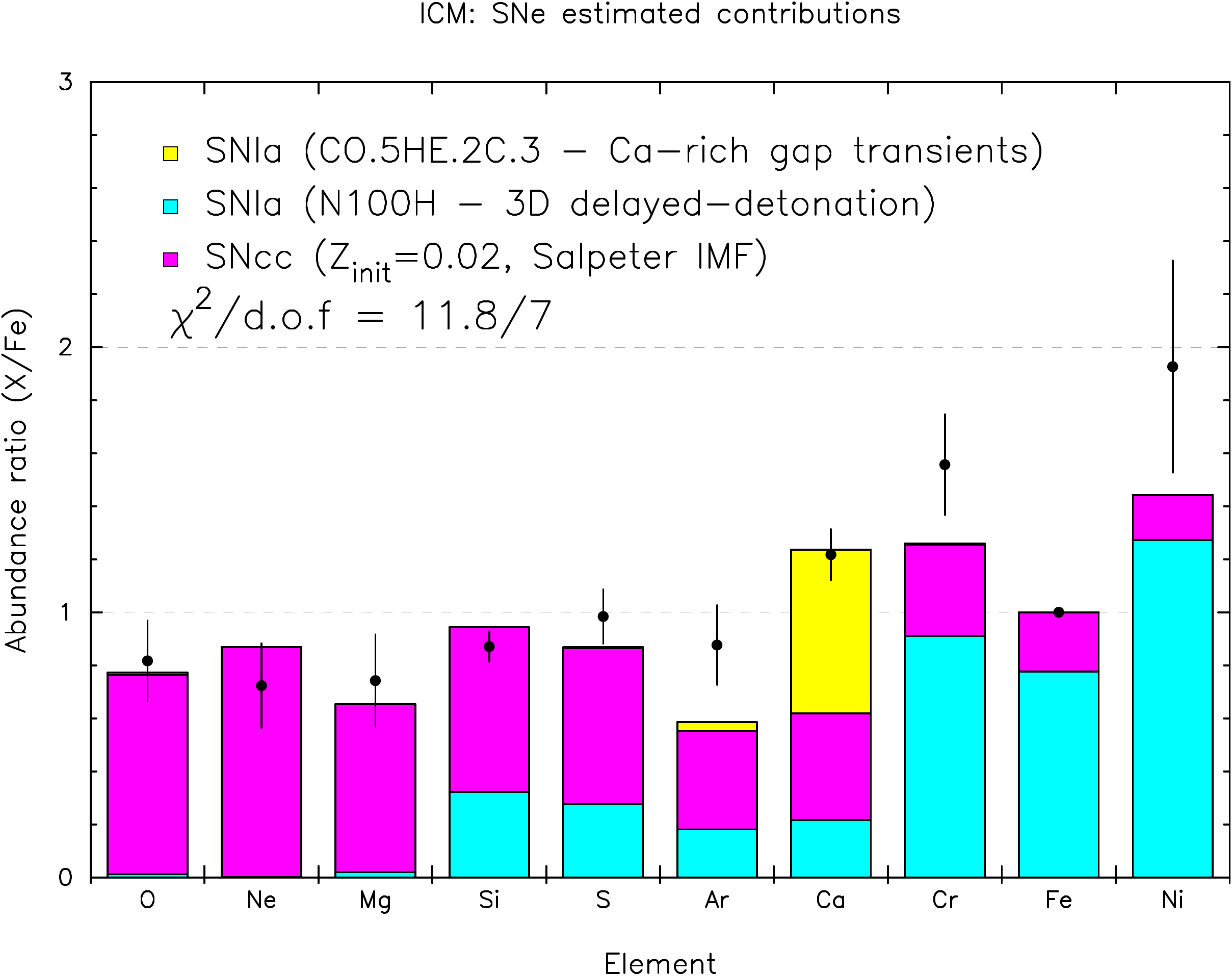} \\

        \caption{\textit{Left}: same as Fig. \ref{fig:SNe_fits_others} (top right), but considering a 2-D SNIa model instead of a Classical SNIa model. \textit{Right}: same as Fig. \ref{fig:SNe_fits_others} (top right), but considering a 3-D SNIa model instead of a Classical SNIa model.}
\label{fig:SNe_fits_2D3D}
\end{figure*}

From Table \ref{table:SNe_fits_2D}, it clearly appears that the use of the 2D models does not improve our fit. In fact, while the (C- and O-) DDT models largely overestimate (>4$\sigma$) the Si/Fe ratio, the C-DEF model overestimates (>2$\sigma$) the Ni/Fe ratio (see also Fig. \ref{fig:SNe_models}, left). Moreover, unlike in Sect. \ref{sect:nickel}, using two 2D SNIa models does not  improve the quality of the fit. The best fit, obtained for the combination Z0.02+O-DDT+CO.5HE.2C.3 ($\chi^2 / \text{d.o.f.} \simeq$ 4.3) is shown in Fig. \ref{fig:SNe_fits_2D3D} (left panel).

 The 3D models (Table \ref{table:SNe_fits_3D}) look somewhat more encouraging. Although the combination Nomoto+3D  clearly does not reproduce the ICM abundance pattern (see Sect. \ref{sect:calcium}), the addition of a Ca-rich gap contribution significantly improves the quality of the fit. In particular, this confirms the Ca/Fe problem discussed earlier, and strengthens the need for an additional contribution, for instance, from Ca-rich gap transients. The favoured SNIa model (N100H) assumes a delayed-detonation explosion, where the core density of the pre-exploding WD is quite high ($5.5 \times 10^9$ g/cm$^3$). We also note that considering two channels of SNIa explosion (as in Sect. \ref{sect:nickel}) does not improve the quality of the fit (Table  \ref{table:SNe_fits_3D}). In fact, the estimated contribution from delayed-detonation SNIa is clearly marginal (typically $\sim$10\% of the total SNIa contribution). The best fit, obtained for the combination Z0.02+N100H+CO.5HE.2C.3 (with a reduced $\chi^2$ of $\sim$1.7) is shown in Fig. \ref{fig:SNe_fits_2D3D} (right panel).

Moreover, although the 3D models agree better with our ICM abundance pattern than the 2D models, we stress that the Classical and/or Bravo (i.e. 1-D) models still  significantly provide the best match to our observations (Table \ref{table:SNe_fits}). This is partly because the multi-dimensional delayed-detonation SNIa models predict a higher Si/Fe ratio than in the 1-D case, making a full compensation by the SNcc yields difficult, since the predicted O/Fe and Si/Fe ratios from SNcc must be rather similar (Fig. \ref{fig:SNe_models}, left). Moreover, the 2-D and 3-D deflagration SNIa models predict systematically lower S/Si and Ar/Si ratios (Fig. \ref{fig:SNe_models}, left), which cannot reproduce our measurements even when accounting for the SNcc contribution.

\subsection{ Mn/Fe ratio}\label{sect:manganese}

In Paper I, we were able to detect Mn in the ICM with >7$\sigma$ (MOS and pn combined), and to constrain an average Mn/Fe abundance under reasonable uncertainties ($\sim$13\%). To our knowledge, this is the first time that the abundance of an odd-Z element has been measured in the ICM. It is commonly known that the bulk of Mn comes from SNIa explosions as SNcc are very inefficient in producing Mn (Fig. \ref{fig:SNe_models}, right). In this section, we discuss two interesting consequences that the observed ICM Mn/Fe ratio (again, witnessing the explosion of billions of SNIa) can have on the SNIa progenitors.

\subsubsection{Metallicity of the SNIa progenitors}\label{sect:SNIa_metallicity}

\begin{figure}[!]
        \centering
                \includegraphics[width=0.49\textwidth]{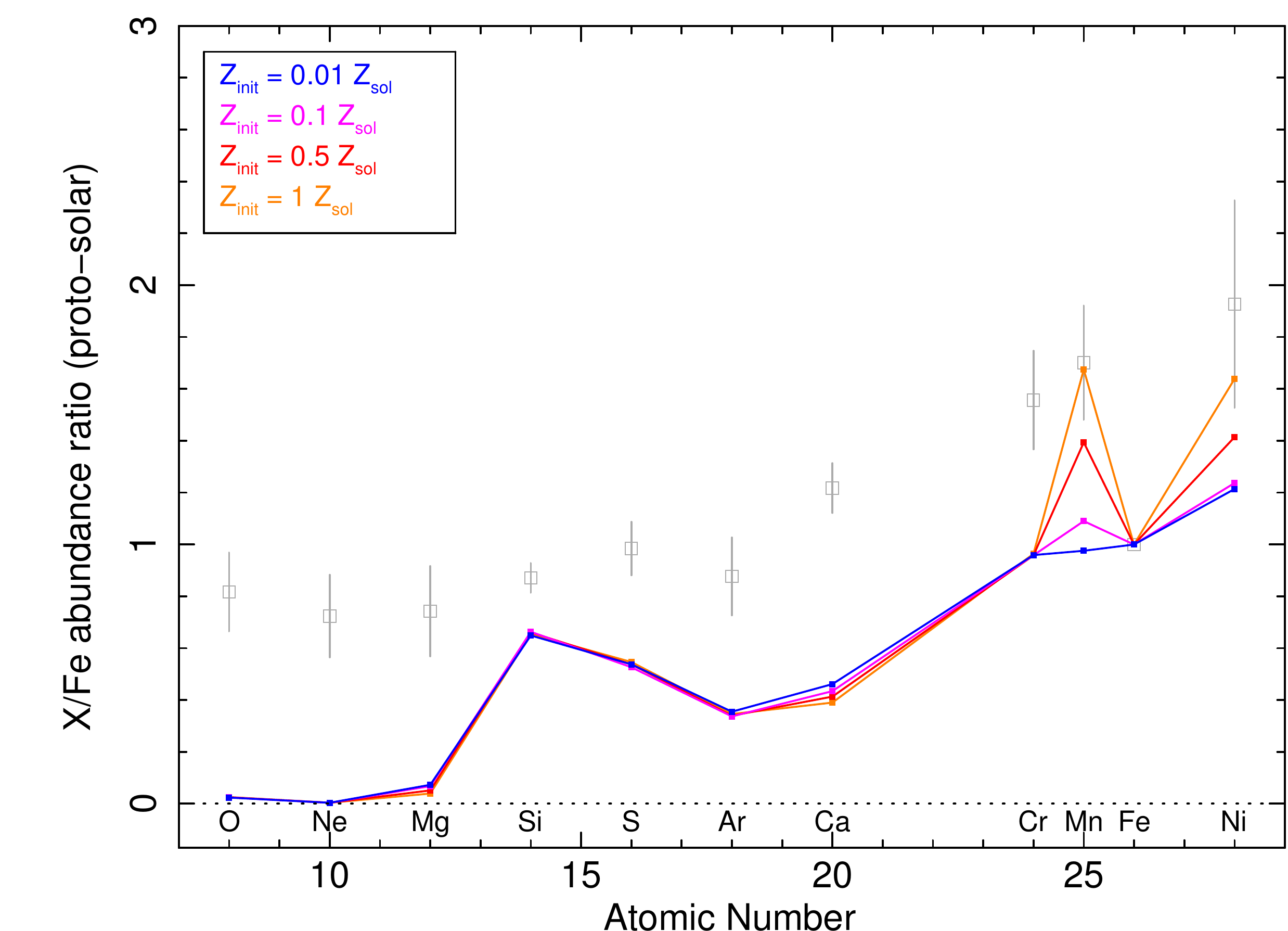}
        \caption{Predicted effects of the initial metallicity of the SNIa progenitor on the X/Fe abundance ratios. The comparison is made using the N100\_Z0.01, N100\_Z0.1, N100\_Z0.5, and N100 3-D delayed-detonation models from \citet{2013MNRAS.429.1156S}. For comparison, the ICM average abundance ratios (inferred from Paper I) are also plotted.}
\label{fig:N100_Zinit}
\end{figure}

\begin{figure}[!]
        \centering
                \includegraphics[width=0.48\textwidth,trim={10 10 10 11},clip]{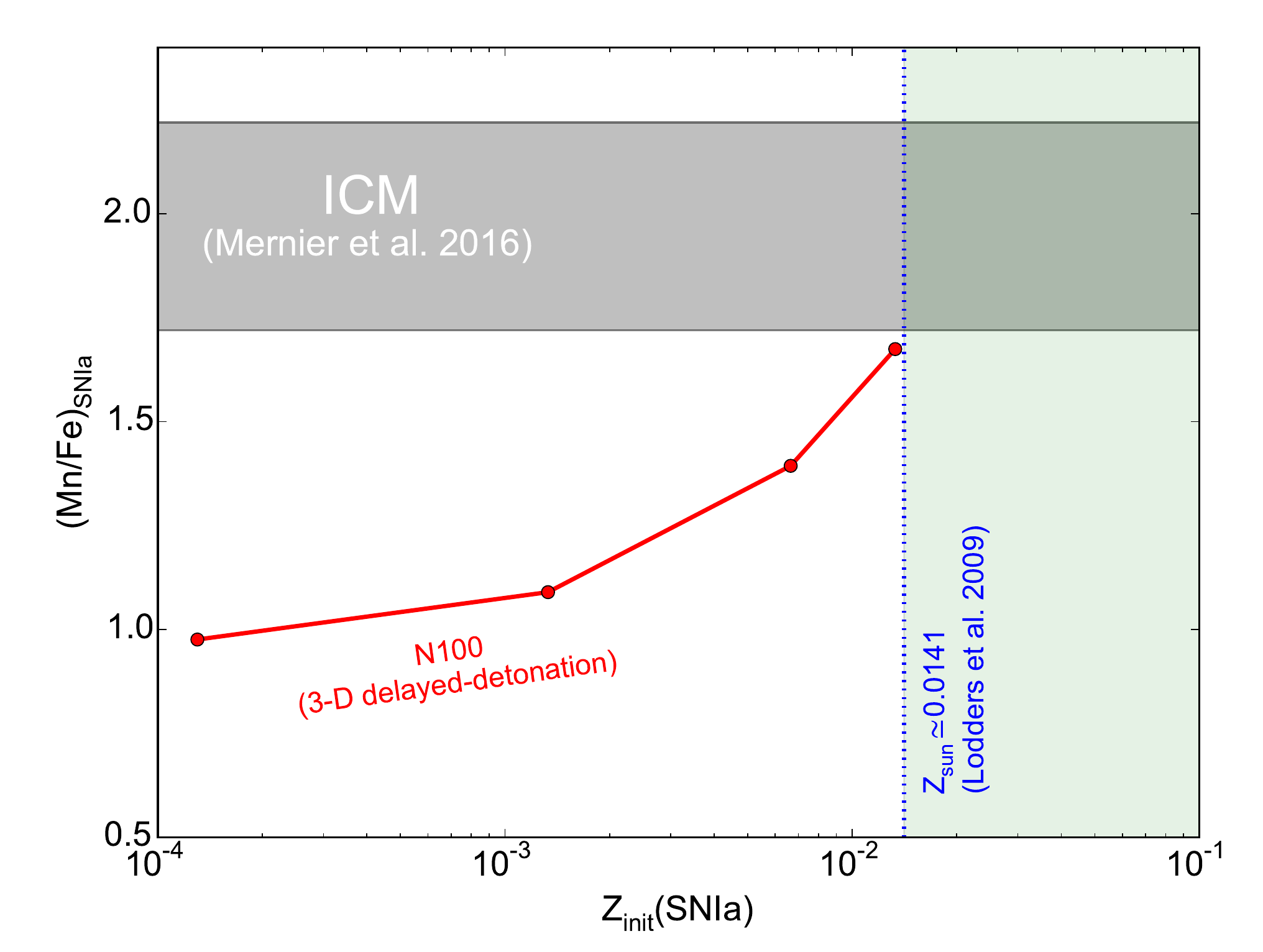}
        \caption{Mn/Fe yields expected from SNIa as a function of the initial metallicity of SNIa progenitors. The red solid line interpolates the estimated Mn/Fe ratio for four  SNIa initial metallicites by \citet{2013MNRAS.429.1156S}, and is compared to the Mn/Fe measurement (expected from SNIa contributions) in the ICM (Paper I and this work) is shown in grey. The dotted blue line shows the solar metallicity \citep{2009LanB...4B...44L}.}
\label{fig:MnFe_ratio}
\end{figure}

\citet{2013MNRAS.429.1156S} calculated the yields from their N100 (3-D delayed-detonation) model, assuming four different initial metallicities ($0.01 Z_\sun$, $0.1 Z_\sun$, $0.5 Z_\sun$, and $1 Z_\sun$) of SNIa progenitors, $Z_\text{init}$(SNIa). Interestingly, the result (Fig. \ref{fig:N100_Zinit}) shows a slight, but clear dependence of the Mn/Fe abundance ratio with $Z_\text{init}$(SNIa) \citep[see also][]{2015MNRAS.447.1484S}. 
Since the bulk of the Mn observed in the ICM is produced by SNIa, we can use our observed Mn/Fe abundance ratio to derive constraints on the average metallicity of the progenitors of SNIa responsible for the enrichment. Following the 1-D yield models best reproducing our abundance pattern, we estimate that respectively $\sim$95\% and $\sim$82\% of the Mn and Fe are produced by SNIa. Taking these factors into account, the average Mn/Fe abundance ratio in the ICM coming from SNIa is $1.97 \pm 0.25$. Again assuming the N100 model for the SNIa contribution, the interpolation of the yields from \citet{2013MNRAS.429.1156S} involves a lower limit of $Z_\text{init}\text{(SNIa)} \gtrsim Z_\sun$ (Fig. \ref{fig:MnFe_ratio}).

The lack of yield models with $Z > 1 Z_\sun$ combined with the uncertainties in our Mn/Fe ICM measurement prevents us from inferring further constraints such as an upper limit to $Z_\text{init}$(SNIa). We also recall that our inferred lower limit depends on the assumed limited Mn production by SNcc. If, for some reason, the Mn production is revised upwards in upcoming SNcc yield models, it would have a strong impact on the inferred limits of the initial metallicities of SNIa progenitors. Moreover, a more complete understanding of the precise relation between $Z_\text{init}$(SNIa) and the Mn yields could  only be achieved by comparing this $Z_\text{init}$(SNIa) dependence in various SNIa yield models. Except N100, no other available deflagration/delayed-detonation model has been calculated for several successive values of $Z_\text{init}$(SNIa). For this reason, we prefer to treat Mn as a peculiar element, and therefore, Mn/Fe was not included in the previous fits (Sect. \ref{sect:even}).

Finally, it is worth noting (see Fig. \ref{fig:N100_Zinit}) that the Ni/Fe ratio from SNIa contributions may also vary with $Z_\text{init}$(SNIa) (at least for metallicities beyond $\sim$0.5 $Z_\sun$). If such a trend is demonstrated in other  delayed-detonation models (e.g. 1-D), a high initial metallicity for SNIa progenitors could be an interesting alternative to the need of an additional deflagration channel, as it might reconcile the high Ni/Fe ratio with the rest of the ICM abundance pattern (Sect. \ref{sect:nickel}). However, a clear relation between Ni/Fe and $Z_\text{init}$ has not yet been established, and the large uncertainties in our measured Ni/Fe ratio do not allow us to explore this possibility further.

\subsubsection{Clues on the nature of SNIa progenitors}\label{sect:SNIa_progenitors}

In principle, the formation channel of the binary system leading to the SNIa explosion (i.e. single-degenerate versus double-degenerate scenario) affects the explosion itself. In the single-degenerate scenario, the explosion occurs during the accretion process from the stellar companion as the total mass of the WD approaches the Chandrasekhar mass limit (near-$M_\text{Ch}$) and leads to the deflagration and/or delayed-detonation explosion mechanisms discussed earlier. In the double-degenerate scenario, when assuming a violent merger of the two WDs (without accretion disc), the explosion is thought to occur well below the Chandrasekhar mass limit (sub-$M_\text{Ch}$) of either WD, and should lead to a pure detonation explosion. Another possibility is that the less massive WD gets disrupted and forms a thick disc that the more massive WD gradually accretes. If the WD rotates rapidly, C may be ignited in its core and lead to a SNIa with a deflagration or a delayed-detonation explosion \citep[e.g.][]{2003ApJ...598.1229P}.

Recently, \citet{2015MNRAS.447.1484S} suggested that the Mn K$\alpha$ line emissivity inferred from the X-ray spectra of SNIa could bring a tight constraint on the scenarios mentioned above as the near-$M_\text{Ch}$ models produce significantly more $^{55}$Fe (later decaying into stable $^{55}$Mn) than the sub-$M_\text{Ch}$ models. On the other hand, \citet{2013A&A...559L...5S} showed that sub-$M_\text{Ch}$ explosion models from \citet[][; violent WD merger with respective masses of 1.1 $M_\sun$ and 0.9 $M_\sun$]{2012ApJ...747L..10P} and \citet[][; violent WD merger with masses of 0.6 $M_\sun$ each]{2013MNRAS.429.1425R} systematically predict sub-solar Mn/Fe abundance ratios, and can hardly explain the proto-solar value of Mn. Although Mn yields from SNIa may be metallicity-dependent (Sect. \ref{sect:SNIa_metallicity}), \citet{2015MNRAS.447.1484S} noted that even the highest-$Z_\text{init}$ (i.e. 1 $Z_\sun$) sub-$M_\text{Ch}$ model produces two times less Mn than the lowest-$Z_\text{init}$ (i.e. 0.01 $Z_\sun$) near-$M_\text{Ch}$ model.

Again assuming that $\sim$95\% of the Mn/Fe ratio measured in the ICM originates from SNIa explosions, our super-proto-solar ICM Mn/Fe ratio constrains this result even more, and suggests that the WD violent merger scenario should be excluded as a dominant SNIa progenitor channel (at least assuming that such a merger produces a pure detonation).

To confirm this claim in a larger context, we re-fit our ICM abundance pattern, this time by including the publicly available yields from the WD violent merger model of \citet{2010Natur.463...61P}. This  sub-$M_\text{Ch}$  
yield is only available for the violent merger of two WDs with equal masses ($M_\text{WD1} = M_\text{WD2} = 0.9$ $M_\sun$; see also Table \ref{table:SNe_models}). We consider two specific cases:
\begin{enumerate}
\item All SNIa (excluding Ca-rich gap transients) originate from violent WD mergers (both of equal mass $M_\text{WD1} = M_\text{WD2} = 0.9$ $M_\sun$), and they occur as sub-$M_\text{Ch}$ explosions (i.e. the Nomoto+sub-$M_\text{Ch}$+Ca-rich gap combination);
\item One part of the SNIa originate from violent WD mergers, the other part originate from another channel, occurring as near-$M_\text{Ch}$ explosions, either deflagration or delayed-detonation (i.e. the Nomoto+sub-$M_\text{Ch}$+3D+Ca-rich gap combination).
\end{enumerate}

In the first case, the fit fails to find any positive contribution for the sub-$M_\text{Ch}$ model. In the second case, the contribution of the sub-$M_\text{Ch}$ SNIa to the enrichment is limited to $\sim$1.3\% of the total number of SNIa, with similar best fits to those we reported in Sect. \ref{sect:calcium} (Nomoto+Classical+Ca-rich combination). This occurs because the Si/Fe ratio predicted by the 0.9\_0.9 model is dramatically higher ($\sim$8) than the observed ratio in the ICM. Again, this favours near-$M_\text{Ch}$ explosions, and discards the violent WD mergers scenario (leading to sub-$M_\text{Ch}$ explosions) as a significant contributor to SNIa nucleosynthesis, at least for the two specific combinations of initial masses discussed above (1.1 $M_\sun$+0.9 $M_\sun$ and 0.9 $M_\sun$+0.9 $M_\sun$). We must note, however, that this result does {not} necessarily discard all the subchannels of the double-degenerate scenario. For instance, as mentioned above, a disruption of the least massive WD, followed by the creation of a thick torus that could feed the most massive WD, may lead to a near-$M_\text{Ch}$ explosion, and to similar yields as used in the Classical, 2D and 3D models. Moreover, we recall that our discussion is entirely based on the current yield predictions. Any substantial change in upcoming yield models of sub-$M_\text{Ch}$ explosions (for instance in the initial masses that are assumed) may potentially challenge  our interpretation.

\subsection{Fraction of low-mass stars that become SNIa}\label{sect:WDtoSNIa}

Since SNcc explosions are the result of the end-of-life of massive stars ($\ge$10 $M_\sun$), the bulk of SNcc events occur very rapidly ($\lesssim$40 Myr) after its associated episode of star formation. On the contrary, SNIa events require a considerable time delay (up to several Gyr), from their zero age low-mass star progenitors to the end of the binary evolution of the corresponding WD(s). 
In our Galaxy, multiple episodes of star formation continuously generate low-mass stars, and make it difficult to directly compare the number of SNIa events and the corresponding number of low-mass stars that have generated them. 

In galaxy clusters, however, the situation is different. In fact, since the bulk of the star formation occurred at the epoch of cluster formation ($z \simeq $ 2--3), and has now dramatically quenched, clusters are an interesting laboratory which allow us to relate the estimated number of low-mass stars to the number of SNIa, and thus estimate the SNIa efficiency, $\eta_\text{Ia}$ (i.e. the fraction of low-mass stars that eventually result in SNIa). Following the approach of \citet{2007A&A...465..345D}, in this section we attempt to estimate $\eta_\text{Ia}$ from the ICM abundance measurements and their best-fit SN yield models.

Quantitatively, assuming a power-law IMF, we can write \citep{2007A&A...465..345D}
\begin{equation}\label{eq:SNIa_to_lowmassstars}
\frac{\text{SNIa}}{\text{SNIa+SNcc}} = \frac{\eta_\text{Ia}\int_{M_\text{low}}^{M_\text{cc}} m^{-(1+x)} \,dm}{\eta_\text{Ia}\int_{M_\text{low}}^{M_\text{cc}} m^{-(1+x)} \,dm + \int_{M_\text{cc}}^{M_\text{up}} m^{-(1+x)} \,dm} ,
\end{equation}
where $M_\text{low}$ and $M_\text{cc}$ are respectively the lower and upper mass limit of stars that eventually result in SNIa, and $M_\text{up}$ is the upper mass limit of massive stars contributing to the ICM enrichment. Low-mass stars are thus comprised between $M_\text{low} < M < M_\text{cc}$ (i.e. $\eta_\text{Ia}$ is estimated for the stars within this range), and massive stars (all producing SNcc) are comprised between $M_\text{cc} < M < M_\text{up}$. 

Based on all our previous fits that are of acceptable quality (i.e. $\chi^2$/d.o.f $\lesssim 2$ in Tables \ref{table:SNe_fits}, \ref{table:SNe_fits_2D}, and \ref{table:SNe_fits_3D}), the SNIa-to-SNe fraction responsible for the enrichment varies within $\sim$29--45\%. The typical $M_\text{low}$ values found in the literature vary between $M_\text{low}=0.9$ $M_\sun$ (the minimum mass allowed for a star to end its life within the Hubble time) and $M_\text{low}=1.5$ $M_\sun$ \citep[to allow the accreting WD to reach a value close to its Chandrasekhar limit within a binary system, e.g.][]{2001ApJ...558..351M,2007A&A...465..345D}. We also assume that the bulk of high-mass stars responsible for the enrichment (i.e. via SNcc explosions) has a non-zero initial metallicity, and therefore we limit $M_\text{up}$ to 50 $M_\sun$. Finally, we allow $M_\text{cc}$ to vary between $\sim$8 $M_\sun$ \citep[e.g.][]{2009ARA&A..47...63S} and $\sim$10 $M_\sun$ \citep[e.g.][]{2013ARA&A..51..457N}. We also assume a Salpeter IMF. From Eq. (\ref{eq:SNIa_to_lowmassstars}), and exploring the different limits reported above, we obtain $\eta_{\text{Ia},0.9} \simeq 1.5$--4\% and $\eta_{\text{Ia},1.5} \simeq 3$--9\% as the fraction of low-mass stars, respectively with $M \ge 0.9$ $M_\sun$ and $M \ge 1.5$ $M_\sun$, that eventually become SNIa. These two estimates are in agreement with previous typical values of 3--10\% reported in the literature \citep[e.g.][]{1996ApJ...462..266Y,2001ApJ...558..351M,2012PASA...29..447M,2013ApJ...773...52L}. Similarly, \citet{2008MNRAS.384..267M} compiled various observational estimates of $\eta_\text{Ia}$ from the literature, this time adopting $M_\text{low}=3$ $M_\sun$, i.e. the most appropriate value for the double-degenerate scenario. In particular, he shows that under this condition the estimate ($\eta_\text{Ia} \simeq 14$--40\%) of \citet{2007A&A...465..345D} brings larger upper limits than the other estimates, always below $\sim$20\% \citep[e.g.][]{2003ApJ...591..749L,2004ApJ...613..189D,2005A&A...433..807M,2006ApJ...648..868S}. In that context, we reconsider our estimate of $\eta_\text{Ia}$, this time by assuming $M_\text{low}=3$ $M_\sun$. We find  $\eta_{\text{Ia},3} \simeq 9$--27\%, hence lowering the maximum estimate of the fraction of low-mass stars that become SNIa.

We recall, however, that we use the instantaneous recycling approximation for such an estimate, and the SN fractions introduced in this work should be interpreted as the fraction of SNIa and SNcc {contributing} to the ICM enrichment. In particular, a higher SNcc lock-up efficiency \citep[i.e. the efficiency for SNcc products to be recycled back into stars instead of enriching the ICM;][]{2013ApJ...773...52L} would make the true SNIa-to-SNe fraction (i.e. accounting for the {total} number of SNIa and SNcc) somewhat lower than our current estimate and, consequently, would lower $\eta_\text{Ia}$ as well. Nevertheless, this consideration requires detailed calculations of stellar and galactic evolutionary models, which is beyond the scope of the present paper.

\subsection{Clues on the metal budget conundrum in clusters}

Previous studies clearly report that, in the ICM of massive ($\gtrsim 10^{13}$--$10^{14}$ $M_\sun$) galaxy clusters, the measured Fe content is far above expectations if we assume the currently favoured SN efficiencies, star formation, IMF, and Fe production rate by SNIa and SNcc \citep[e.g.][]{1993ApJ...419...52R,2006ApJ...648..230L,2013ApJ...773...52L,2014MNRAS.444.3581R,2016arXiv160304858Y}. This conundrum on the metal budget in galaxy clusters has not yet been solved. In this section, we explore two possibilities that have been proposed in the literature, and that directly depend on the ICM abundance pattern we report in this work.

\subsubsection{Effect of the IMF on core-collapse yields}\label{sect:IMFs}

One of the possible solutions to this conundrum would be a completely different IMF in galaxy clusters from that measured in the field. In particular, if low-metallicity environments favour formation of higher mass stars, invoking a top-heavy (i.e. flat, $x=-1$) IMF could potentially boost the Fe production by SNcc and reconcile the Fe stellar production and the Fe mass in the ICM on clusters scales \citep[e.g.][]{2005MNRAS.358.1247N}.

\begin{figure}[!]
        \centering
                \includegraphics[width=0.49\textwidth]{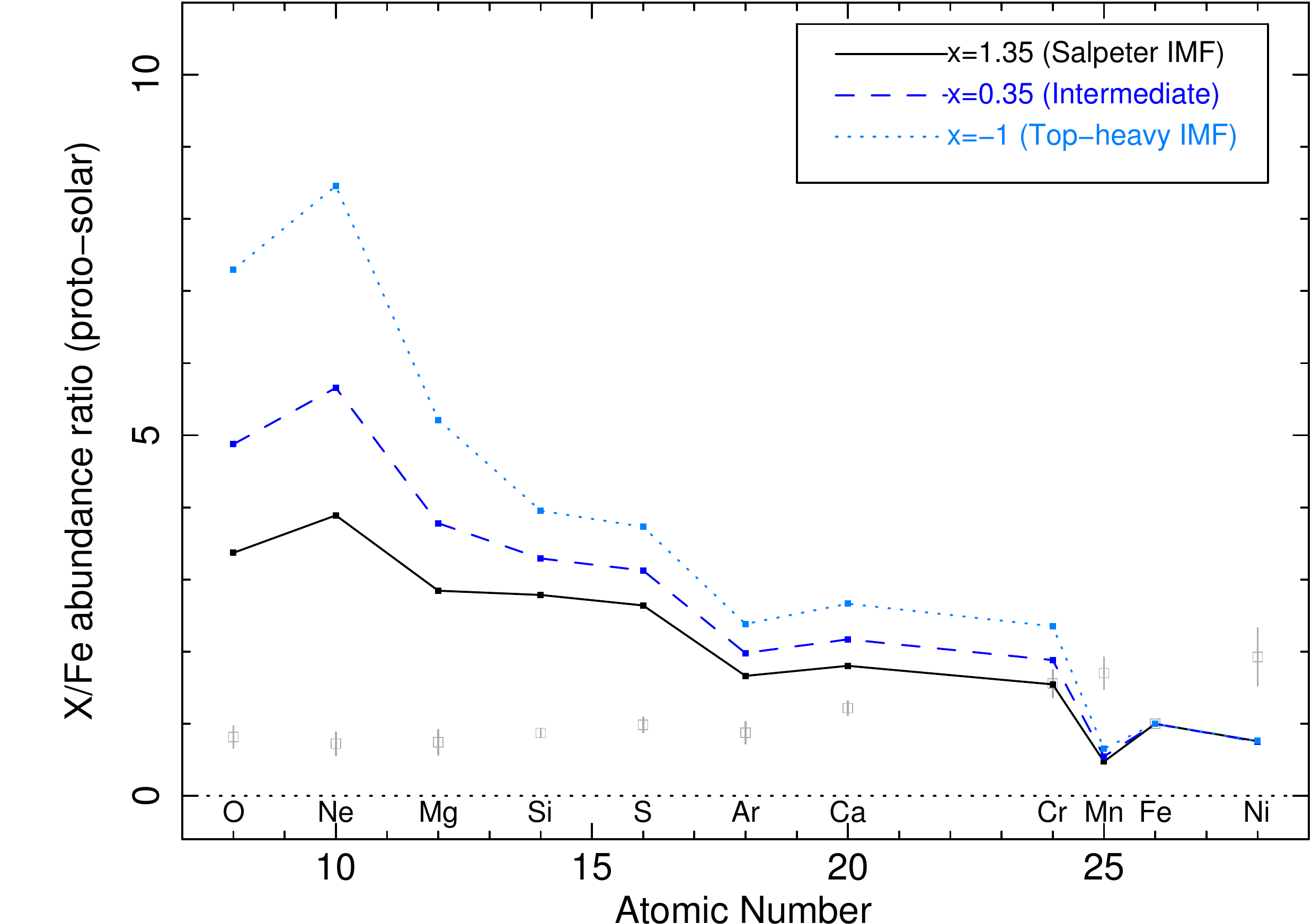}
        \caption{Predicted X/Fe abundances from the  Z0.02 SNcc yield model (Nomoto), computed for three different IMFs (Salpeter IMF, intermediate case, and top-heavy IMF). For comparison, the ICM average abundance ratios (inferred from Paper I) are also plotted.}
\label{fig:IMFs}
\end{figure}

We explore this possibility by fitting the same combinations of SN models discussed above to our measured ICM abundance pattern, this time integrating the SNcc yields over a top-heavy IMF. As can be seen in Fig. \ref{fig:IMFs}, the slope of the IMF has an effect on the relative abundance of all the $\alpha$-elements, in particular on the Ne/Mg ratio. Assuming a top-heavy IMF, the Nomoto+Classical case gives slightly more acceptable results, improving the  best-fit reduced $\chi^2$ from $\sim$2.8 (Z0.008+WDD2, Salpeter IMF) to $\sim$2.6 (Z0.008+WDD2, top-heavy IMF). In all other cases, however, these best fits are either comparable to or less acceptable than when assuming a Salpeter IMF. In other words, despite our effort in constraining the X/Fe abundance uncertainties, the large error bars of O/Fe, Ne/Fe, and Mg/Fe prevents us from deriving any firm conclusion on the IMF in galaxy clusters/groups.

\subsubsection{Contribution from pair-instability supernovae?}\label{sect:PISNe}

\begin{figure}[!]
        \centering
                \includegraphics[width=0.49\textwidth]{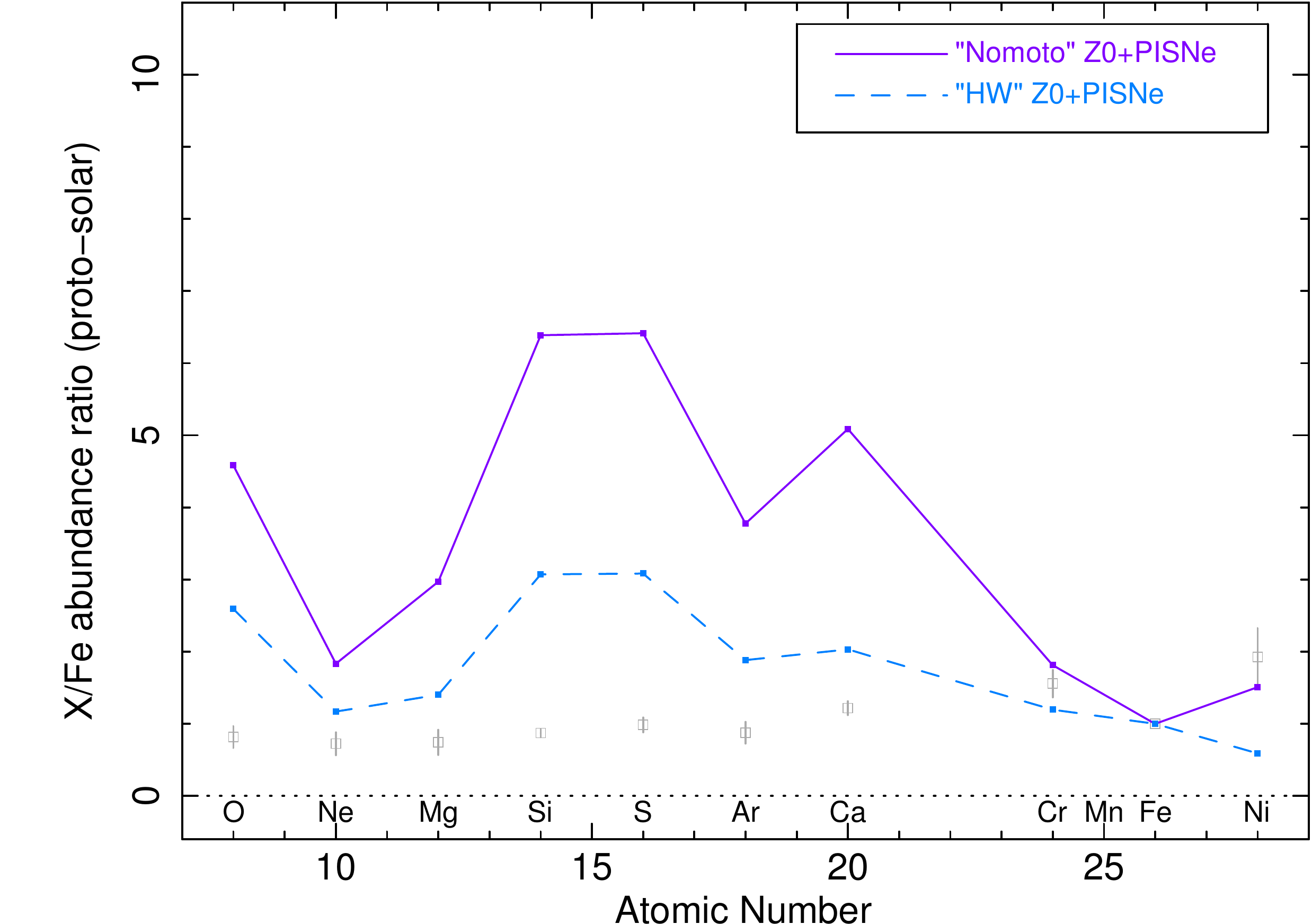}
        \caption{Predicted X/Fe abundances from two $Z_\text{init}$=0 SNcc models with an additional yields contribution from PISNe (Nomoto and HW Z0+PISN models, see also Table \ref{table:SNe_models}). For comparison, the ICM average abundance ratios (inferred from Paper I) are also plotted.}
\label{fig:PISNe}
\end{figure}

As an alternative to a different IMF in cluster galaxies, \citet{2014MNRAS.441.2134M} suggest that the large Fe content found in the ICM may be explained by accounting for the contribution of pair-instability supernovae (PISNe) to the overall enrichment. In fact, by convention, the IMF is often restricted to an upper limit of $\sim$40 $M_\sun$ or $\sim$140 $M_\sun$ (depending on the assumed $Z_\text{init}$ for SNcc), whereas PISNe (typically estimated to occur between 140--300 $M_\sun$) are thought to produce much larger amounts of metals than SNcc or SNIa. To explore this possibility, we redo the same abundance fits as described above, this time by incorporating nucleosynthetic yields of PISNe, and by extending the upper mass limit of the Salpeter IMF to the largest mass for which PISNe can produce and eject metals. We assume that only stars with $Z_\text{init} = 0$ can give rise to PISNe \citep{2013ARA&A..51..457N}. Two distinct models (see Table \ref{table:SNe_models} and Fig. \ref{fig:PISNe}) are considered here.
\begin{enumerate}
\item The Nomoto Z0+PISNe model: the Z0 model presented earlier (up to 140 $M_\sun$) from \citet{2013ARA&A..51..457N}, combined with the PISNe model (140--300 $M_\sun$) from \citet{2002ApJ...565..385U}.
\item The "HW" Z0+PISNe model: the $Z_\text{init}$=0 model for SNcc (up to 100 $M_\sun$, where we assume equal contributions from models with SNcc energies of 0.3, 0.6, 0.9, 1.2, 1.5, 2.4, and $3.0 \times 10^{51}$ erg) from \citet{2010ApJ...724..341H}, combined with the PISNe model (140--260 $M_\sun$) from \citet{2002ApJ...567..532H}. This model has also been considered in order to remain consistent with the analysis of \citet{2014MNRAS.441.2134M}.
\end{enumerate}

When using these extended models instead of the regular SNcc models used so far, we find that the fits are always significantly poorer than previously reported. In particular, the O/Ne and O/Mg ratios, as well as the Si/Fe ratios (and sometimes S/Fe), are dramatically overestimated by the models. This strongly suggests that a contribution from PISNe to the enrichment is unlikely (or insufficient) to explain the large amount of metals found in the ICM. Nevertheless, as mentioned earlier, such a claim is dependent on the model yields proposed so far.


\section{Enrichment in the solar neighbourhood}\label{sect:proto-solar}


In addition to the ICM average abundance pattern presented in Paper I and this work, the X/Fe abundance ratios from our solar system (Fig. \ref{fig:cluster_vs_solar}) offers an interesting additional dataset to test predictions from SN yield models. In particular, it is  reasonable to assume that the SNIa explosion channel(s) enriching galaxy clusters and the solar neighbourhood must be the same, presumably with a different relative fraction of SNIa and SNcc having contributed to the enrichment. This potentially brings an additional constraint on the specific SNIa explosion models to favour. However, the SNcc progenitors that enriched the Milky Way and the ICM did not necessarily have the same average initial metallicity. Consequently, the various sets of SN yield models presented in this paper should be fitted separately to the ICM abundance pattern (Sect. \ref{sect:ICM}) on the one hand, and to the proto-solar values (this section) on the other hand.

In the following we always assume a Salpeter IMF. Similarly to Sect. \ref{sect:even}, we ignore the proto-solar Mn/Fe ratio because of its possible dependence on the metallicity of SNIa progenitors, which itself depends on the considered SNIa model (Sect. \ref{sect:SNIa_progenitors}). We also note that a significant part of the nitrogen, fluorine, and sodium yields is thought to be produced by AGB stars, which we do not consider in this work. Therefore, in the following we also ignore the proto-solar N/Fe, F/Fe, and Na/Fe ratios.

\begin{figure}[!]
        \centering
                \includegraphics[width=0.49\textwidth]{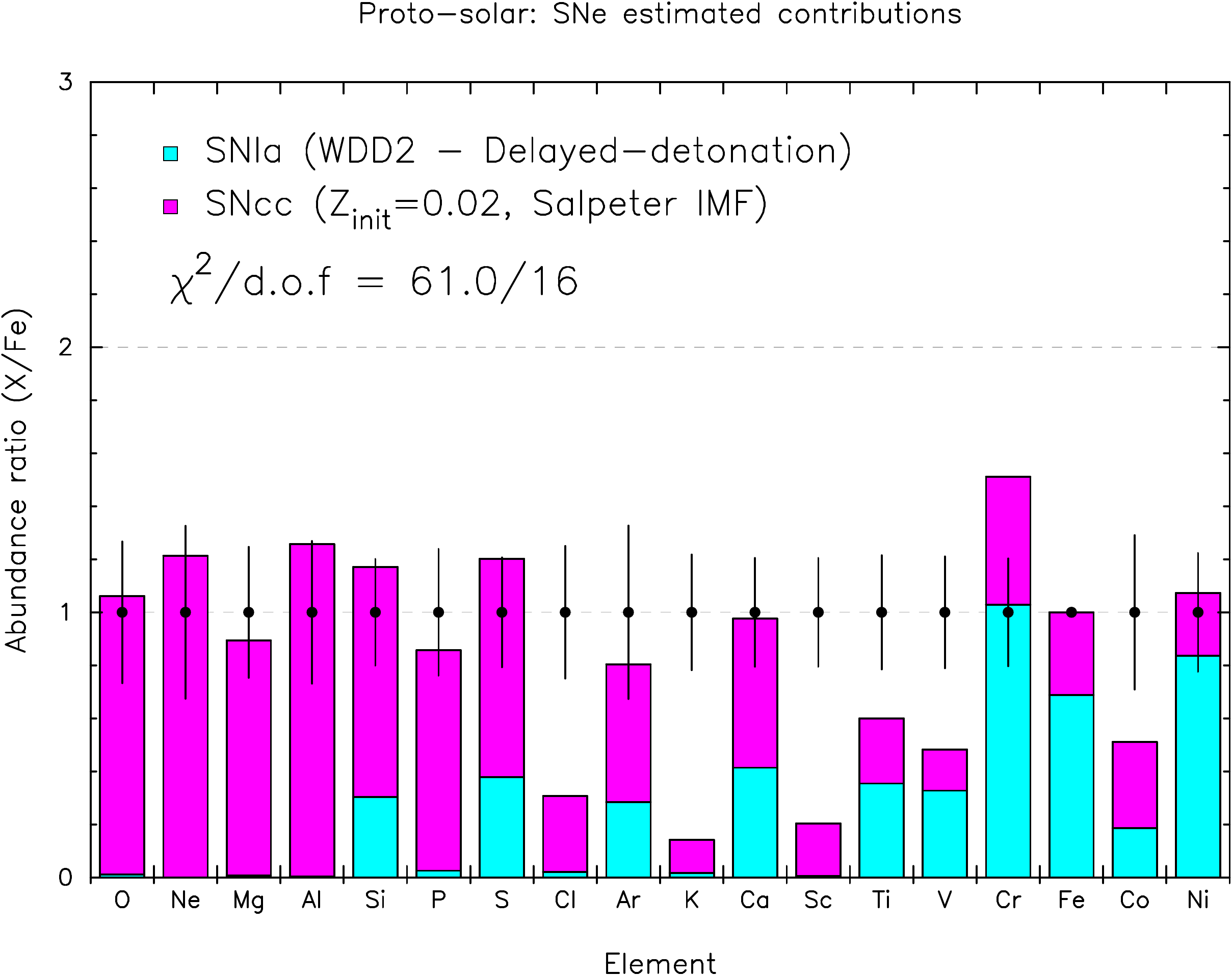}

        \caption{Abundance ratios versus atomic numbers in the proto-solar abundance pattern \citep{2009LanB...4B...44L}. The histograms show the yields contribution of a best-fit combination of one Classical SNIa model (WDD2) and one Nomoto SNcc ($Z_\text{init} = 0.02$, and Salpeter IMF) model.}
\label{fig:SNe_fits_sol}
\end{figure}

We start by considering sets of one SNcc and one SNIa model, namely the Nomoto+Classical, Nomoto+Bravo, Nomoto+2D, and Nomoto+3D combinations. The five best fits of each combination are listed in Table \ref{table:SNe_fits_sol}. With a reduced $\chi^2$ of $\sim$3.8, the best fit is obtained for a combination Z0.02+WDD2, and is shown in Fig. \ref{fig:SNe_fits_sol} (left). Clearly, these sets of models do not  reproduce well the proto-solar abundance pattern. The main reason is that the ratios of Cl/Fe, K/Fe, Sc/Fe, Ti/Fe, V/Fe, and Co/Fe are systematically underestimated (with >2$\sigma$, >3$\sigma$, >3$\sigma$, >1$\sigma$, >2$\sigma$, and >1$\sigma$, respectively) by all the models. In some cases, the Cr/Fe ratio is somewhat overestimated. Such discrepancies have already been  reported in the literature by using Galactic evolution models \citep[e.g.][]{2006ApJ...653.1145K,2010A&A...522A..32R,2011MNRAS.414.3231K,2013ARA&A..51..457N}. Although the problem has not yet been discussed  in detail, it is possible that the $\nu$-process significantly increases the production of these elements in SNcc \citep{2011MNRAS.414.3231K,2013ARA&A..51..457N}. Except for these specific cases, the ratios of the other elements (mostly even-Z)  are correctly reproduced. For comparison, if we include only the X/Fe ratios of the even-Z elements that could be measured in the ICM, we find that the best fit is obtained for a Z0.02+WDD3 combination, with a reduced $\chi^2$ of $\sim$0.6, and a SN fraction of $\sim$20\%. Based on Table \ref{table:SNe_fits_sol}, some additional remarks are worth mentioning.

\begin{table}[!]
\begin{centering}
\caption{Results of various SN fits to the proto-solar abundance ratios adapted from \citet{2009LanB...4B...44L}. Only one SNIa and one SNcc model have been fitted, and for each case we only show the five best fits (sorted by increasing $\chi^2 / \text{d.o.f.}$).}             
\label{table:SNe_fits_sol}
\begin{tabular}{c c c c}        
\hline \hline                

SNcc & SNIa & $\frac{\text{SNIa}}{\text{SNIa}+\text{SNcc}}$ & $\chi^2 / \text{d.o.f.}$ \\
 & & \\
\hline
Nomoto & Classical &   &   \\    
\hline
Z0.02 & WDD2  & $0.20$ &  $61.0/16$ \\
Z0.02 & CDD2  & $0.19$ &  $61.8/16$ \\
Z0.02 & WDD3  & $0.18$ &  $62.7/16$ \\
Z0.02 & W70  & $0.19$ &  $72.3/16$ \\
Z0.008 & WDD3  & $0.18$ &  $75.6/16$ \\
\hline
Nomoto & Bravo &   &  \\    
\hline
Z0.02 & DDTc  & $0.20$ &  $64.5/16$ \\
Z0.02 & DDTa  & $0.15$ &  $66.2/16$ \\
Z0.008 & DDTa & $0.15$ &  $77.6/16$ \\
Z0.008 & DDTc & $0.21$ &  $80.0/16$ \\
Z0.004 & DDTa  & $0.16$ &  $94.2/16$ \\
\hline
Nomoto & 2D &   &  \\    
\hline
Z0.02 & O-DDT & $0.24$ &  $67.8/16$ \\
Z0.008 & O-DDT & $0.25$ &  $81.4/16$ \\
Z0.004 & O-DDT & $0.27$ &  $97.1/16$ \\
Z0.001 & O-DDT & $0.30$ &  $99.5/16$ \\
Z0\_cut & O-DDT & $0.28$ &  $111.5/16$ \\
\hline
Nomoto & 3D &   &  \\    
\hline
Z0.02 & N100H & $0.18$ &  $62.0/16$ \\
Z0.02 & N40  & $0.20$ &  $63.9/16$ \\
Z0.02 & N100  & $0.21$ &  $64.1/16$ \\
Z0.02 & N20 & $0.17$ &  $64.4/16$ \\
Z0.02 & N150 & $0.22$ &  $64.6/16$ \\

\hline                                   
\end{tabular}
\par\end{centering}
\end{table}

First,  all the best fits are reached for a SNcc initial metallicity $Z_\text{init} = 0.02$. In fact, the Z0.02 model is clearly favoured by the the O/Fe, Ne/Fe, Mg/Fe, and Al/Fe abundance ratios, whose elements are almost entirely produced by SNcc. Of course, an enrichment of the solar system with SNcc already having solar initial metallicities is not possible. On the other hand, $Z_\text{init} = 0.008$ may be on the low side for the average contributing SNcc, and no model assuming intermediate values of $Z_\text{init}$ is available so far. Therefore, Z0.02 is the most suitable model, but this statement must be carfeully interpreted.
Because of the poor quality of the fits, it is more difficult to favour one specific SNIa model. However, it appears that the delayed-detonation models are systematically preferred to the deflagration models. The preference of the WDD2, DDTc, O-DDT, and N100H models, respectively in the Classical, Bravo, 2D, and 3D categories, is also consistent with the best fits of the ICM abundance pattern (Tables \ref{table:SNe_fits}, \ref{table:SNe_fits_2D}, and \ref{table:SNe_fits_3D}).

Second,  the proto-solar abundance pattern does not need an additional contribution from Ca-rich gap transients as the Ca/Fe ratio is already successfully reproduced. Such a result may not be surprising if, as already discussed in Sect. \ref{sect:calcium}, Ca-rich gap SNe explode preferably in the galaxy outskirts, hence easily enriching the ICM, while their contribution in enriching the solar neighbourhood should be quite limited.

Similarly, an additional SNIa component (to account for the possible diversity of SNIa explosions, Sect. \ref{sect:nickel}) does not improve the quality of the fits. Quantitatively, when fitting an additional SNIa model to the proto-solar abundance pattern, the contribution of delayed-detonation SNIa to the local enrichment is systematically $\gtrsim$10 times more important than any additional contribution of deflagration SNIa.

Finally, the estimated enriching SNIa-to-SNe fraction is systematically lower for the enrichment of the solar neighbourhood ($\sim$15--30\%) than for the ICM enrichment. Here again, this is not surprising. If the bulk of local SNIa progenitors result from a recent star formation, most of them had not yet exploded at the epoch of the formation of the Sun, and could not have contributed to the enrichment of the solar neighbourhood (on the contrary of SNcc progenitors, which explode shortly after their formation). On the other hand, in galaxy clusters, almost all potential SNIa have exploded (except perhaps those with extremely long delay times) giving rise to a substantial fraction of SNIa yields. We cannot exclude, however, that other effects (e.g. different lock-up efficiencies) may also play a role to explain the difference between local and cluster SNIa-to-SNe fractions.

We must emphasise that our approach in comparing the SN yield models to the proto-solar abundance ratios is purely empirical as we are just interested in a direct comparison between local and ICM enrichments. Ideally, a full Galactic enrichment study would require a complete evolutionary model (taking account of star formation, infall, outflows, Galactic age, binary populations, etc.) to be compared to the abundances in stars in the solar neighbourhood \citep[e.g.][]{2006ApJ...653.1145K,2009ApJ...707.1466K,2010A&A...522A..32R,2011MNRAS.414.3231K}. However, such a detailed study is beyond the scope of this present work.

\section{Summary and conclusions}\label{sect:conclusion}

In this paper, we have made use of the most precise and complete average X/Fe abundance ratios measured in the ICM so far (derived in Paper I), in order to constrain properties of SNIa, SNcc, and their relative contribution to the enrichment at the scale of galaxy clusters. Our main results can be summarised as follows.

\begin{itemize}

\item Whereas a simple combination of one Classical SNIa and one Nomoto SNcc model is sufficient to explain most of the X/Fe abundance ratios in the solar neighbourhood, this is clearly not the case in the ICM. In particular, this set of models cannot explain the high Ca/Fe and Ni/Fe ratios found in the ICM. In other words, the ICM seems to be particularly Ca- and Ni-rich.

\item The Ca/Fe ratio can be successfully reproduced if we assume a significant contribution to the enrichment from Ca-rich gap transients, a recently discovered class of SNe which explode as WDs and are surprisingly rich in Ca. Based on the available models, a significant mixing ($\sim$30\% in mass)  between the C-O core and the He layer of the pre-exploding WD is necessary to reconcile the enriching fraction of Ca-rich gap transients to the rates inferred from optical observations (less than $\sim$10\% of the total number of SNIa events). However, a higher Ca-rich SNe contribution to the enrichment (this time assuming no mixing of the WD material) cannot be excluded  as these objects preferentially explode far away from the galactic centres, and their yields could thus be easily mixed into the ICM, as compared to the Galaxy or solar neighbourhood. This could also explain why no significant Ca-rich SNe contribution is necessary in the enrichment of the solar neighbourhood.

Alternatively to this scenario, the Ca/Fe predicted ratio can be reconciled with our measurement by using a SNIa delayed-detonation model based on the Tycho supernova remnant \citep{2006ApJ...645.1373B}. Unfortunately, the uncertainties in our measurements do not allow us to favour one of these two scenarios.

\item The best way to successfully reproduce the Ni/Fe ratio in the ICM (simultaneously with the other X/Fe ratios) is to invoke a diversity in SNIa explosions, with $\sim$50--77\% of deflagration SNIa, and the remaining fraction of delayed-detonation SNIa. On the other hand, the proto-solar abundance pattern does not require such a diversity in SNIa, and clearly favours the delayed-detonation explosion as the dominant channel ($\gtrsim$90\% of SNIa).

\item The Mn/Fe ratio---measured in the ICM for the first time---can in principle bring useful constraints on the initial metallicity of SNIa progenitors. Assuming a limited ($\sim$5\%) Mn production from SNcc, we find that $Z_\text{init}\text{(SNIa)} \gtrsim 1 Z_\sun$. This result is, of course, very dependent on the assumed yields, and more SNIa models (with varying values of $Z_\text{init}$(SNIa)) are clearly needed to extend the discussion further. The initial metallicity of SNIa progenitors also affects the Ni production, and could be considered as a possible alternative to the co-existence of both delayed-detonation and deflagration SNIa explosions.

In addition to this consideration, the high Mn/Fe ratio suggests that a negligible contribution from a hypothetical sub-$M_\text{Ch}$ SNIa channel (associated with a detonative explosion). Considering the models available so far, this could imply that the majority of SNIa contributing to the ICM (and Galactic) enrichment were {not} produced by violent WD mergers.

\item Interestingly, the recent 2-D \citep{2010ApJ...712..624M} and 3-D \citep{2013MNRAS.429.1156S,2014MNRAS.438.1762F} SNIa models are less efficient in reproducing the ICM (and proto-solar) abundance pattern than the basic 1-D \citep{1999ApJS..125..439I,2006ApJ...645.1373B} SNIa models. In particular, the multi-dimensional models tend to overproduce Si, whereas the Si/Fe ratio in the ICM is very well constrained by our observations.

\item Based on all the models that reasonably reproduce our ICM abundance pattern, we estimate that $\sim$29--45\% of the SNe contributing to the enrichment are SNIa, the remaining part coming from SNcc. This fraction is systematically higher than in the solar neighbourhood ($\sim$15--25\%), and could be explained by the rapid quenching of star formation in galaxy clusters shortly after their assembling. Such a SNIa fraction in the ICM also implies, under rough assumptions, that the fraction of low-mass stars that become SNIa ranges between 1.5\% and 27\% (depending on the assumed lower mass limit), in agreement with most of the observations.

\item The uncertainties in the ICM abundance ratios prevent us from putting tight constraints on the IMF, the initial parameters of the deflagration/delayed-detonation WD explosion leading to SNIa, or the initial metallicity of SNcc progenitors. For the latter, however, our measurements can reasonably exclude a SNcc enrichment with a zero initial metallicity, meaning that the SNcc progenitors that enriched the ICM must have been previously pre-enriched. Similarly, a significant enrichment of the ICM by PISNe (in addition to SNcc) can reasonably be excluded.

\end{itemize}

\subsection{Future directions}\label{sect:future}

As we have seen throughout this paper, the determination of several SNIa and SNcc properties (as well as their relative contribution to the ICM enrichment) can, in principle, be constrained from the ICM (and proto-solar) abundance pattern. Although we showed that some combinations of models and hypotheses can be ruled out with a high degree of certainty, it is still impossible to clearly favour one specific combination of SN models. For instance, as we have shown in Sect. \ref{sect:nickel}, if one wants to confirm (or rule out) the bimodality in SNIa explosions that enrich the ICM, a very precise determination of the Ni/Fe ratio is essential, but currently not possible. Similarly, although the amount of metals released by SNcc is in principle sensitive to the the shape of the IMF, we cannot clearly favour one specific IMF with our current results and the available models (Sect. \ref{sect:IMFs}). Finally, despite our detailed discussion on the possible contribution of Ca-rich gap transients to the enrichment, the high Ca/Fe ratio measured in the ICM remains an open issue.
In order to better constrain the stellar origins of the ICM enrichment, further improvements on many aspects are clearly needed.

First, both SNIa and SNcc yield models still suffer from uncertainties (e.g. see discussion in Appendix \ref{sect:ecapture}). Major improvements of these models (in particular, better agreements in the yields of physically comparable models) are thus crucial for the purpose of this study.
Second, an ongoing effort should be made to reduce the systematic uncertainties in the solar and meteoritic abundances, as, together with ICM abundances, they may provide further constraints to SNIa and SNcc yield models. 
Third, future direct studies of SNe will be complementary to this work. In particular, improvements in estimating the relative fraction of SNIa, and particularly of Ca-rich gap transients, exploding in the Galaxy and in galaxy clusters could bring additional valuable constraints on the SN models to favour. Conversely, the estimates from our study may be useful to complement future direct observations of SNe.

Finally, the uncertainties in the ICM abundances must be reduced as well. For instance, the discrepancies between atomic data have been greatly reduced over the past decades, but the atomic codes should be continuously updated. Similarly, calibration issues in the current X-ray instruments have been improved, but still largely contribute to the current uncertainties \citep[e.g.][Paper I]{2015A&A...575A..30S,2015A&A...575A..37M}. In Paper I we showed that adding more cluster data would  not reduce the current uncertainties in the ICM abundance pattern. Therefore, next-generation X-ray missions (in particular using micro-calorimeter arrays, which should significantly improve the spectral resolution currently achieved with CCDs) are crucial to provide a better general understanding of the ICM enrichment and the origin of metals in the Universe.

\begin{acknowledgements}
We thank the anonymous referee for useful comments which helped to improve the paper. This work is partly based on the \textit{XMM-Newton} AO-12 proposal ``\emph{The XMM-Newton view of chemical enrichment in bright galaxy clusters and groups}'' (PI: de Plaa), and is a part of the CHEERS (CHEmical Evolution Rgs cluster Sample) collaboration. The authors thank its members, as well as Liyi Gu and Craig Sarazin for helpful discussions. P.K. thanks Steve Allen and Ondrej Urban for support and hospitality at Stanford University. Y.Y.Z. acknowledges support from the German BMWI through the Verbundforschung under grant 50\,OR\,1506. This work is based on observations obtained with \textit{XMM-Newton}, an ESA science mission with instruments and contributions directly funded by ESA member states and the USA (NASA). The SRON Netherlands Institute for Space Research is supported financially by NWO, the Netherlands Organisation for Space Research.
\end{acknowledgements}

\bibliography{Core_abundances-II_hk}{}

\begin{thebibliography}{110}
\expandafter\ifx\csname natexlab\endcsname\relax\def\natexlab#1{#1}\fi

\bibitem[{{Arnett}(1973)}]{1973ARA&A..11...73A}
{Arnett}, W.~D. 1973, \araa, 11, 73

\bibitem[{{Badenes} {et~al.}(2006){Badenes}, {Borkowski}, {Hughes}, {Hwang}, \&
  {Bravo}}]{2006ApJ...645.1373B}
{Badenes}, C., {Borkowski}, K.~J., {Hughes}, J.~P., {Hwang}, U., \& {Bravo}, E.
  2006, \apj, 645, 1373

\bibitem[{{Badenes} {et~al.}(2003){Badenes}, {Bravo}, {Borkowski}, \&
  {Dom{\'{\i}}nguez}}]{2003ApJ...593..358B}
{Badenes}, C., {Bravo}, E., {Borkowski}, K.~J., \& {Dom{\'{\i}}nguez}, I. 2003,
  \apj, 593, 358

\bibitem[{{B{\"o}hringer} {et~al.}(2005){B{\"o}hringer}, {Matsushita},
  {Finoguenov}, {Xue}, \& {Churazov}}]{2005AdSpR..36..677B}
{B{\"o}hringer}, H., {Matsushita}, K., {Finoguenov}, A., {Xue}, Y., \&
  {Churazov}, E. 2005, Advances in Space Research, 36, 677

\bibitem[{{B{\"o}hringer} \& {Werner}(2010)}]{2010A&ARv..18..127B}
{B{\"o}hringer}, H. \& {Werner}, N. 2010, \aapr, 18, 127

\bibitem[{{Brachwitz} {et~al.}(2000){Brachwitz}, {Dean}, {Hix}, {Iwamoto},
  {Langanke}, {Mart{\'{\i}}nez-Pinedo}, {Nomoto}, {Strayer}, {Thielemann}, \&
  {Umeda}}]{2000ApJ...536..934B}
{Brachwitz}, F., {Dean}, D.~J., {Hix}, W.~R., {et~al.} 2000, \apj, 536, 934

\bibitem[{{Branch}(2001)}]{2001PASP..113..169B}
{Branch}, D. 2001, \pasp, 113, 169

\bibitem[{{Branch} {et~al.}(2004){Branch}, {Thomas}, {Baron}, {Kasen},
  {Hatano}, {Nomoto}, {Filippenko}, {Li}, \& {Rudy}}]{2004ApJ...606..413B}
{Branch}, D., {Thomas}, R.~C., {Baron}, E., {et~al.} 2004, \apj, 606, 413

\bibitem[{{Bravo} {et~al.}(1996){Bravo}, {Tornambe}, {Dominguez}, \&
  {Isern}}]{1996A&A...306..811B}
{Bravo}, E., {Tornambe}, A., {Dominguez}, I., \& {Isern}, J. 1996, \aap, 306,
  811

\bibitem[{{Burbidge} {et~al.}(1957){Burbidge}, {Burbidge}, {Fowler}, \&
  {Hoyle}}]{1957RvMP...29..547B}
{Burbidge}, E.~M., {Burbidge}, G.~R., {Fowler}, W.~A., \& {Hoyle}, F. 1957,
  Reviews of Modern Physics, 29, 547

\bibitem[{{Cameron}(1957)}]{1957AJ.....62....9C}
{Cameron}, A.~G.~W. 1957, \aj, 62, 9

\bibitem[{{Cao} {et~al.}(2015){Cao}, {Kulkarni}, {Howell}, {Gal-Yam},
  {Kasliwal}, {Valenti}, {Johansson}, {Amanullah}, {Goobar}, {Sollerman},
  {Taddia}, {Horesh}, {Sagiv}, {Cenko}, {Nugent}, {Arcavi}, {Surace},
  {Wo{\'z}niak}, {Moody}, {Rebbapragada}, {Bue}, \&
  {Gehrels}}]{2015Natur.521..328C}
{Cao}, Y., {Kulkarni}, S.~R., {Howell}, D.~A., {et~al.} 2015, \nat, 521, 328

\bibitem[{{Crosby} {et~al.}(2013){Crosby}, {O'Shea}, {Peruta}, {Beers}, \&
  {Tumlinson}}]{2013arXiv1312.0606C}
{Crosby}, B.~D., {O'Shea}, B.~W., {Peruta}, C., {Beers}, T.~C., \& {Tumlinson},
  J. 2013, ArXiv e-prints

\bibitem[{{Dahlen} {et~al.}(2004){Dahlen}, {Strolger}, {Riess}, {Mobasher},
  {Chary}, {Conselice}, {Ferguson}, {Fruchter}, {Giavalisco}, {Livio}, {Madau},
  {Panagia}, \& {Tonry}}]{2004ApJ...613..189D}
{Dahlen}, T., {Strolger}, L.-G., {Riess}, A.~G., {et~al.} 2004, \apj, 613, 189

\bibitem[{{De Grandi} \& {Molendi}(2009)}]{2009A&A...508..565D}
{De Grandi}, S. \& {Molendi}, S. 2009, \aap, 508, 565

\bibitem[{{de Plaa} {et~al.}(2007){de Plaa}, {Werner}, {Bleeker}, {Vink},
  {Kaastra}, \& {M{\'e}ndez}}]{2007A&A...465..345D}
{de Plaa}, J., {Werner}, N., {Bleeker}, J.~A.~M., {et~al.} 2007, \aap, 465, 345

\bibitem[{{de Plaa} {et~al.}(2006){de Plaa}, {Werner}, {Bykov}, {Kaastra},
  {M{\'e}ndez}, {Vink}, {Bleeker}, {Bonamente}, \&
  {Peterson}}]{2006A&A...452..397D}
{de Plaa}, J., {Werner}, N., {Bykov}, A.~M., {et~al.} 2006, \aap, 452, 397

\bibitem[{{Dupke} \& {White}(2000)}]{2000ApJ...528..139D}
{Dupke}, R.~A. \& {White}, III, R.~E. 2000, \apj, 528, 139

\bibitem[{{Dutton} {et~al.}(2012){Dutton}, {Mendel}, \&
  {Simard}}]{2012MNRAS.422L..33D}
{Dutton}, A.~A., {Mendel}, J.~T., \& {Simard}, L. 2012, \mnras, 422, 33

\bibitem[{{Filippenko} {et~al.}(2003){Filippenko}, {Chornock}, {Swift},
  {Modjaz}, {Simcoe}, \& {Rauch}}]{2003IAUC.8159....2F}
{Filippenko}, A.~V., {Chornock}, R., {Swift}, B., {et~al.} 2003, \iaucirc,
  8159, 2

\bibitem[{{Fink} {et~al.}(2014){Fink}, {Kromer}, {Seitenzahl},
  {Ciaraldi-Schoolmann}, {R{\"o}pke}, {Sim}, {Pakmor}, {Ruiter}, \&
  {Hillebrandt}}]{2014MNRAS.438.1762F}
{Fink}, M., {Kromer}, M., {Seitenzahl}, I.~R., {et~al.} 2014, \mnras, 438, 1762

\bibitem[{{Finoguenov} {et~al.}(2002){Finoguenov}, {Matsushita},
  {B{\"o}hringer}, {Ikebe}, \& {Arnaud}}]{2002A&A...381...21F}
{Finoguenov}, A., {Matsushita}, K., {B{\"o}hringer}, H., {Ikebe}, Y., \&
  {Arnaud}, M. 2002, \aap, 381, 21

\bibitem[{{Foley}(2015)}]{2015MNRAS.452.2463F}
{Foley}, R.~J. 2015, \mnras, 452, 2463

\bibitem[{{Fuller} {et~al.}(1982){Fuller}, {Fowler}, \&
  {Newman}}]{1982ApJS...48..279F}
{Fuller}, G.~M., {Fowler}, W.~A., \& {Newman}, M.~J. 1982, \apjs, 48, 279

\bibitem[{{Gamezo} {et~al.}(2005){Gamezo}, {Khokhlov}, \&
  {Oran}}]{2005ApJ...623..337G}
{Gamezo}, V.~N., {Khokhlov}, A.~M., \& {Oran}, E.~S. 2005, \apj, 623, 337

\bibitem[{{Hamuy} {et~al.}(2000){Hamuy}, {Trager}, {Pinto}, {Phillips},
  {Schommer}, {Ivanov}, \& {Suntzeff}}]{2000AJ....120.1479H}
{Hamuy}, M., {Trager}, S.~C., {Pinto}, P.~A., {et~al.} 2000, \aj, 120, 1479

\bibitem[{{Hatano} {et~al.}(2000){Hatano}, {Branch}, {Lentz}, {Baron},
  {Filippenko}, \& {Garnavich}}]{2000ApJ...543L..49H}
{Hatano}, K., {Branch}, D., {Lentz}, E.~J., {et~al.} 2000, \apjl, 543, L49

\bibitem[{{Heger} \& {Woosley}(2002)}]{2002ApJ...567..532H}
{Heger}, A. \& {Woosley}, S.~E. 2002, \apj, 567, 532

\bibitem[{{Heger} \& {Woosley}(2010)}]{2010ApJ...724..341H}
{Heger}, A. \& {Woosley}, S.~E. 2010, \apj, 724, 341

\bibitem[{{Hillebrandt} {et~al.}(2013){Hillebrandt}, {Kromer}, {R{\"o}pke}, \&
  {Ruiter}}]{2013FrPhy...8..116H}
{Hillebrandt}, W., {Kromer}, M., {R{\"o}pke}, F.~K., \& {Ruiter}, A.~J. 2013,
  Frontiers of Physics, 8, 116

\bibitem[{{Hillebrandt} \& {Niemeyer}(2000)}]{2000ARA&A..38..191H}
{Hillebrandt}, W. \& {Niemeyer}, J.~C. 2000, \araa, 38, 191

\bibitem[{{Hopkins} \& {Beacom}(2006)}]{2006ApJ...651..142H}
{Hopkins}, A.~M. \& {Beacom}, J.~F. 2006, \apj, 651, 142

\bibitem[{{Howell}(2011)}]{2011NatCo...2E.350H}
{Howell}, D.~A. 2011, Nature Communications, 2, 350

\bibitem[{{Iben} \& {Tutukov}(1984)}]{1984ApJS...54..335I}
{Iben}, Jr., I. \& {Tutukov}, A.~V. 1984, \apjs, 54, 335

\bibitem[{{Iwamoto} {et~al.}(1999){Iwamoto}, {Brachwitz}, {Nomoto},
  {Kishimoto}, {Umeda}, {Hix}, \& {Thielemann}}]{1999ApJS..125..439I}
{Iwamoto}, K., {Brachwitz}, F., {Nomoto}, K., {et~al.} 1999, \apjs, 125, 439

\bibitem[{{Jerkstrand} {et~al.}(2015){Jerkstrand}, {Timmes}, {Magkotsios},
  {Sim}, {Fransson}, {Spyromilio}, {M{\"u}ller}, {Heger}, {Sollerman}, \&
  {Smartt}}]{2015ApJ...807..110J}
{Jerkstrand}, A., {Timmes}, F.~X., {Magkotsios}, G., {et~al.} 2015, \apj, 807,
  110

\bibitem[{{Jha} {et~al.}(2006){Jha}, {Branch}, {Chornock}, {Foley}, {Li},
  {Swift}, {Casebeer}, \& {Filippenko}}]{2006AJ....132..189J}
{Jha}, S., {Branch}, D., {Chornock}, R., {et~al.} 2006, \aj, 132, 189

\bibitem[{{Kaastra} {et~al.}(1996){Kaastra}, {Mewe}, \&
  {Nieuwenhuijzen}}]{1996uxsa.conf..411K}
{Kaastra}, J.~S., {Mewe}, R., \& {Nieuwenhuijzen}, H. 1996, in UV and X-ray
  Spectroscopy of Astrophysical and Laboratory Plasmas, ed. K.~{Yamashita} \&
  T.~{Watanabe}, 411--414

\bibitem[{{Kasliwal} {et~al.}(2012){Kasliwal}, {Kulkarni}, {Gal-Yam}, {Nugent},
  {Sullivan}, {Bildsten}, {Yaron}, {Perets}, {Arcavi}, {Ben-Ami}, {Bhalerao},
  {Bloom}, {Cenko}, {Filippenko}, {Frail}, {Ganeshalingam}, {Horesh}, {Howell},
  {Law}, {Leonard}, {Li}, {Ofek}, {Polishook}, {Poznanski}, {Quimby},
  {Silverman}, {Sternberg}, \& {Xu}}]{2012ApJ...755..161K}
{Kasliwal}, M.~M., {Kulkarni}, S.~R., {Gal-Yam}, A., {et~al.} 2012, \apj, 755,
  161

\bibitem[{{Khokhlov}(1989)}]{1989MNRAS.239..785K}
{Khokhlov}, A.~M. 1989, \mnras, 239, 785

\bibitem[{{Khokhlov}(1991)}]{1991A&A...245L..25K}
{Khokhlov}, A.~M. 1991, \aap, 245, L25

\bibitem[{{Kobayashi} {et~al.}(2011){Kobayashi}, {Karakas}, \&
  {Umeda}}]{2011MNRAS.414.3231K}
{Kobayashi}, C., {Karakas}, A.~I., \& {Umeda}, H. 2011, \mnras, 414, 3231

\bibitem[{{Kobayashi} \& {Nomoto}(2009)}]{2009ApJ...707.1466K}
{Kobayashi}, C. \& {Nomoto}, K. 2009, \apj, 707, 1466

\bibitem[{{Kobayashi} {et~al.}(2006){Kobayashi}, {Umeda}, {Nomoto}, {Tominaga},
  \& {Ohkubo}}]{2006ApJ...653.1145K}
{Kobayashi}, C., {Umeda}, H., {Nomoto}, K., {Tominaga}, N., \& {Ohkubo}, T.
  2006, \apj, 653, 1145

\bibitem[{{Kromer} {et~al.}(2013){Kromer}, {Fink}, {Stanishev}, {Taubenberger},
  {Ciaraldi-Schoolman}, {Pakmor}, {R{\"o}pke}, {Ruiter}, {Seitenzahl}, {Sim},
  {Blanc}, {Elias-Rosa}, \& {Hillebrandt}}]{2013MNRAS.429.2287K}
{Kromer}, M., {Fink}, M., {Stanishev}, V., {et~al.} 2013, \mnras, 429, 2287

\bibitem[{{Langanke} \& {Martinez-Pinedo}(1998)}]{1998PhLB..436...19L}
{Langanke}, K. \& {Martinez-Pinedo}, G. 1998, Physics Letters B, 436, 19

\bibitem[{{Langanke} \& {Mart{\'{\i}}nez-Pinedo}(2001)}]{2001ADNDT..79....1L}
{Langanke}, K. \& {Mart{\'{\i}}nez-Pinedo}, G. 2001, Atomic Data and Nuclear
  Data Tables, 79, 1

\bibitem[{{Li} {et~al.}(2011){Li}, {Chornock}, {Leaman}, {Filippenko},
  {Poznanski}, {Wang}, {Ganeshalingam}, \& {Mannucci}}]{2011MNRAS.412.1473L}
{Li}, W., {Chornock}, R., {Leaman}, J., {et~al.} 2011, \mnras, 412, 1473

\bibitem[{{Li} {et~al.}(2001){Li}, {Filippenko}, {Treffers}, {Riess}, {Hu}, \&
  {Qiu}}]{2001ApJ...546..734L}
{Li}, W., {Filippenko}, A.~V., {Treffers}, R.~R., {et~al.} 2001, \apj, 546, 734

\bibitem[{{Lin} {et~al.}(2003){Lin}, {Mohr}, \&
  {Stanford}}]{2003ApJ...591..749L}
{Lin}, Y.-T., {Mohr}, J.~J., \& {Stanford}, S.~A. 2003, \apj, 591, 749

\bibitem[{{Lodders} {et~al.}(2009){Lodders}, {Palme}, \&
  {Gail}}]{2009LanB...4B...44L}
{Lodders}, K., {Palme}, H., \& {Gail}, H.-P. 2009, Landolt B{\"o}rnstein, 44

\bibitem[{{Loewenstein}(2006)}]{2006ApJ...648..230L}
{Loewenstein}, M. 2006, \apj, 648, 230

\bibitem[{{Loewenstein}(2013)}]{2013ApJ...773...52L}
{Loewenstein}, M. 2013, \apj, 773, 52

\bibitem[{{Madau} \& {Dickinson}(2014)}]{2014ARA&A..52..415M}
{Madau}, P. \& {Dickinson}, M. 2014, \araa, 52, 415

\bibitem[{{Maeda} {et~al.}(2010){Maeda}, {R{\"o}pke}, {Fink}, {Hillebrandt},
  {Travaglio}, \& {Thielemann}}]{2010ApJ...712..624M}
{Maeda}, K., {R{\"o}pke}, F.~K., {Fink}, M., {et~al.} 2010, \apj, 712, 624

\bibitem[{{Mannucci} {et~al.}(2006){Mannucci}, {Della Valle}, \&
  {Panagia}}]{2006MNRAS.370..773M}
{Mannucci}, F., {Della Valle}, M., \& {Panagia}, N. 2006, \mnras, 370, 773

\bibitem[{{Mannucci} {et~al.}(2005){Mannucci}, {Della Valle}, {Panagia},
  {Cappellaro}, {Cresci}, {Maiolino}, {Petrosian}, \&
  {Turatto}}]{2005A&A...433..807M}
{Mannucci}, F., {Della Valle}, M., {Panagia}, N., {et~al.} 2005, \aap, 433, 807

\bibitem[{{Maoz}(2008)}]{2008MNRAS.384..267M}
{Maoz}, D. 2008, \mnras, 384, 267

\bibitem[{{Maoz} \& {Mannucci}(2012)}]{2012PASA...29..447M}
{Maoz}, D. \& {Mannucci}, F. 2012, \pasa, 29, 447

\bibitem[{{Maoz} {et~al.}(2014){Maoz}, {Mannucci}, \&
  {Nelemans}}]{2014ARA&A..52..107M}
{Maoz}, D., {Mannucci}, F., \& {Nelemans}, G. 2014, \araa, 52, 107

\bibitem[{{Matteucci} \& {Chiappini}(2005)}]{2005PASA...22...49M}
{Matteucci}, F. \& {Chiappini}, C. 2005, \pasa, 22, 49

\bibitem[{{Matteucci} \& {Recchi}(2001)}]{2001ApJ...558..351M}
{Matteucci}, F. \& {Recchi}, S. 2001, \apj, 558, 351

\bibitem[{{Mernier} {et~al.}(2015){Mernier}, {de Plaa}, {Lovisari}, {Pinto},
  {Zhang}, {Kaastra}, {Werner}, \& {Simionescu}}]{2015A&A...575A..37M}
{Mernier}, F., {de Plaa}, J., {Lovisari}, L., {et~al.} 2015, \aap, 575, A37

\bibitem[{{Mernier} {et~al.}(2016){Mernier}, {de Plaa}, {Pinto}, {Kaastra},
  {Kosec}, {Zhang}, {Mao}, \& {Werner}}]{2016arXiv160601165M}
{Mernier}, F., {de Plaa}, J., {Pinto}, C., {et~al.} 2016, ArXiv e-prints
  [arXiv:1606.01165]

\bibitem[{{Morsony} {et~al.}(2014){Morsony}, {Heath}, \&
  {Workman}}]{2014MNRAS.441.2134M}
{Morsony}, B.~J., {Heath}, C., \& {Workman}, J.~C. 2014, \mnras, 441, 2134

\bibitem[{{Mulchaey} {et~al.}(2014){Mulchaey}, {Kasliwal}, \&
  {Kollmeier}}]{2014ApJ...780L..34M}
{Mulchaey}, J.~S., {Kasliwal}, M.~M., \& {Kollmeier}, J.~A. 2014, \apjl, 780,
  L34

\bibitem[{{Nagashima} {et~al.}(2005){Nagashima}, {Lacey}, {Baugh}, {Frenk}, \&
  {Cole}}]{2005MNRAS.358.1247N}
{Nagashima}, M., {Lacey}, C.~G., {Baugh}, C.~M., {Frenk}, C.~S., \& {Cole}, S.
  2005, \mnras, 358, 1247

\bibitem[{{Niemeyer} \& {Woosley}(1997)}]{1997ApJ...475..740N}
{Niemeyer}, J.~C. \& {Woosley}, S.~E. 1997, \apj, 475, 740

\bibitem[{{Nomoto} {et~al.}(2013){Nomoto}, {Kobayashi}, \&
  {Tominaga}}]{2013ARA&A..51..457N}
{Nomoto}, K., {Kobayashi}, C., \& {Tominaga}, N. 2013, \araa, 51, 457

\bibitem[{{Nomoto} {et~al.}(2006){Nomoto}, {Tominaga}, {Umeda}, {Kobayashi}, \&
  {Maeda}}]{2006NuPhA.777..424N}
{Nomoto}, K., {Tominaga}, N., {Umeda}, H., {Kobayashi}, C., \& {Maeda}, K.
  2006, Nuclear Physics A, 777, 424

\bibitem[{{Olling} {et~al.}(2015){Olling}, {Mushotzky}, {Shaya}, {Rest},
  {Garnavich}, {Tucker}, {Kasen}, {Margheim}, \&
  {Filippenko}}]{2015Natur.521..332O}
{Olling}, R.~P., {Mushotzky}, R., {Shaya}, E.~J., {et~al.} 2015, \nat, 521, 332

\bibitem[{{Pakmor} {et~al.}(2010){Pakmor}, {Kromer}, {R{\"o}pke}, {Sim},
  {Ruiter}, \& {Hillebrandt}}]{2010Natur.463...61P}
{Pakmor}, R., {Kromer}, M., {R{\"o}pke}, F.~K., {et~al.} 2010, \nat, 463, 61

\bibitem[{{Pakmor} {et~al.}(2012){Pakmor}, {Kromer}, {Taubenberger}, {Sim},
  {R{\"o}pke}, \& {Hillebrandt}}]{2012ApJ...747L..10P}
{Pakmor}, R., {Kromer}, M., {Taubenberger}, S., {et~al.} 2012, \apjl, 747, L10

\bibitem[{{Perets} {et~al.}(2011){Perets}, {Gal-yam}, {Crockett}, {Anderson},
  {James}, {Sullivan}, {Neill}, \& {Leonard}}]{2011ApJ...728L..36P}
{Perets}, H.~B., {Gal-yam}, A., {Crockett}, R.~M., {et~al.} 2011, \apjl, 728,
  L36

\bibitem[{{Perets} {et~al.}(2010){Perets}, {Gal-Yam}, {Mazzali}, {Arnett},
  {Kagan}, {Filippenko}, {Li}, {Arcavi}, {Cenko}, {Fox}, {Leonard}, {Moon},
  {Sand}, {Soderberg}, {Anderson}, {James}, {Foley}, {Ganeshalingam}, {Ofek},
  {Bildsten}, {Nelemans}, {Shen}, {Weinberg}, {Metzger}, {Piro}, {Quataert},
  {Kiewe}, \& {Poznanski}}]{2010Natur.465..322P}
{Perets}, H.~B., {Gal-Yam}, A., {Mazzali}, P.~A., {et~al.} 2010, \nat, 465, 322

\bibitem[{{Phillips} {et~al.}(2007){Phillips}, {Li}, {Frieman}, {Blinnikov},
  {DePoy}, {Prieto}, {Milne}, {Contreras}, {Folatelli}, {Morrell}, {Hamuy},
  {Suntzeff}, {Roth}, {Gonz{\'a}lez}, {Krzeminski}, {Filippenko}, {Freedman},
  {Chornock}, {Jha}, {Madore}, {Persson}, {Burns}, {Wyatt}, {Murphy}, {Foley},
  {Ganeshalingam}, {Serduke}, {Krisciunas}, {Bassett}, {Becker}, {Dilday},
  {Eastman}, {Garnavich}, {Holtzman}, {Kessler}, {Lampeitl}, {Marriner},
  {Frank}, {Marshall}, {Miknaitis}, {Sako}, {Schneider}, {van der Heyden}, \&
  {Yasuda}}]{2007PASP..119..360P}
{Phillips}, M.~M., {Li}, W., {Frieman}, J.~A., {et~al.} 2007, \pasp, 119, 360

\bibitem[{{Piersanti} {et~al.}(2003){Piersanti}, {Gagliardi}, {Iben}, \&
  {Tornamb{\'e}}}]{2003ApJ...598.1229P}
{Piersanti}, L., {Gagliardi}, S., {Iben}, Jr., I., \& {Tornamb{\'e}}, A. 2003,
  \apj, 598, 1229

\bibitem[{{Pinto} {et~al.}(2015){Pinto}, {Sanders}, {Werner}, {de Plaa},
  {Fabian}, {Zhang}, {Kaastra}, {Finoguenov}, \&
  {Ahoranta}}]{2015A&A...575A..38P}
{Pinto}, C., {Sanders}, J.~S., {Werner}, N., {et~al.} 2015, \aap, 575, A38

\bibitem[{{Renzini} \& {Andreon}(2014)}]{2014MNRAS.444.3581R}
{Renzini}, A. \& {Andreon}, S. 2014, \mnras, 444, 3581

\bibitem[{{Renzini} {et~al.}(1993){Renzini}, {Ciotti}, {D'Ercole}, \&
  {Pellegrini}}]{1993ApJ...419...52R}
{Renzini}, A., {Ciotti}, L., {D'Ercole}, A., \& {Pellegrini}, S. 1993, \apj,
  419, 52

\bibitem[{{Riess} {et~al.}(1998){Riess}, {Filippenko}, {Challis},
  {Clocchiatti}, {Diercks}, {Garnavich}, {Gilliland}, {Hogan}, {Jha},
  {Kirshner}, {Leibundgut}, {Phillips}, {Reiss}, {Schmidt}, {Schommer},
  {Smith}, {Spyromilio}, {Stubbs}, {Suntzeff}, \&
  {Tonry}}]{1998AJ....116.1009R}
{Riess}, A.~G., {Filippenko}, A.~V., {Challis}, P., {et~al.} 1998, \aj, 116,
  1009

\bibitem[{{Romano} {et~al.}(2010){Romano}, {Karakas}, {Tosi}, \&
  {Matteucci}}]{2010A&A...522A..32R}
{Romano}, D., {Karakas}, A.~I., {Tosi}, M., \& {Matteucci}, F. 2010, \aap, 522,
  A32

\bibitem[{{Ruiter} {et~al.}(2013){Ruiter}, {Sim}, {Pakmor}, {Kromer},
  {Seitenzahl}, {Belczynski}, {Fink}, {Herzog}, {Hillebrandt}, {R{\"o}pke}, \&
  {Taubenberger}}]{2013MNRAS.429.1425R}
{Ruiter}, A.~J., {Sim}, S.~A., {Pakmor}, R., {et~al.} 2013, \mnras, 429, 1425

\bibitem[{{Sato} {et~al.}(2007){Sato}, {Tokoi}, {Matsushita}, {Ishisaki},
  {Yamasaki}, {Ishida}, \& {Ohashi}}]{2007ApJ...667L..41S}
{Sato}, K., {Tokoi}, K., {Matsushita}, K., {et~al.} 2007, \apjl, 667, L41

\bibitem[{{Scalzo} {et~al.}(2014){Scalzo}, {Ruiter}, \&
  {Sim}}]{2014MNRAS.445.2535S}
{Scalzo}, R.~A., {Ruiter}, A.~J., \& {Sim}, S.~A. 2014, \mnras, 445, 2535

\bibitem[{{Schellenberger} {et~al.}(2015){Schellenberger}, {Reiprich},
  {Lovisari}, {Nevalainen}, \& {David}}]{2015A&A...575A..30S}
{Schellenberger}, G., {Reiprich}, T.~H., {Lovisari}, L., {Nevalainen}, J., \&
  {David}, L. 2015, \aap, 575, A30

\bibitem[{{Schmidt} {et~al.}(1998){Schmidt}, {Suntzeff}, {Phillips},
  {Schommer}, {Clocchiatti}, {Kirshner}, {Garnavich}, {Challis}, {Leibundgut},
  {Spyromilio}, {Riess}, {Filippenko}, {Hamuy}, {Smith}, {Hogan}, {Stubbs},
  {Diercks}, {Reiss}, {Gilliland}, {Tonry}, {Maza}, {Dressler}, {Walsh}, \&
  {Ciardullo}}]{1998ApJ...507...46S}
{Schmidt}, B.~P., {Suntzeff}, N.~B., {Phillips}, M.~M., {et~al.} 1998, \apj,
  507, 46

\bibitem[{{Seitenzahl} {et~al.}(2013{\natexlab{a}}){Seitenzahl}, {Cescutti},
  {R{\"o}pke}, {Ruiter}, \& {Pakmor}}]{2013A&A...559L...5S}
{Seitenzahl}, I.~R., {Cescutti}, G., {R{\"o}pke}, F.~K., {Ruiter}, A.~J., \&
  {Pakmor}, R. 2013{\natexlab{a}}, \aap, 559, L5

\bibitem[{{Seitenzahl} {et~al.}(2013{\natexlab{b}}){Seitenzahl},
  {Ciaraldi-Schoolmann}, {R{\"o}pke}, {Fink}, {Hillebrandt}, {Kromer},
  {Pakmor}, {Ruiter}, {Sim}, \& {Taubenberger}}]{2013MNRAS.429.1156S}
{Seitenzahl}, I.~R., {Ciaraldi-Schoolmann}, F., {R{\"o}pke}, F.~K., {et~al.}
  2013{\natexlab{b}}, \mnras, 429, 1156

\bibitem[{{Seitenzahl} {et~al.}(2015){Seitenzahl}, {Summa}, {Krau{\ss}}, {Sim},
  {Diehl}, {Els{\"a}sser}, {Fink}, {Hillebrandt}, {Kromer}, {Maeda},
  {Mannheim}, {Pakmor}, {R{\"o}pke}, {Ruiter}, \&
  {Wilms}}]{2015MNRAS.447.1484S}
{Seitenzahl}, I.~R., {Summa}, A., {Krau{\ss}}, F., {et~al.} 2015, \mnras, 447,
  1484

\bibitem[{{Simionescu} {et~al.}(2009){Simionescu}, {Werner}, {B{\"o}hringer},
  {Kaastra}, {Finoguenov}, {Br{\"u}ggen}, \& {Nulsen}}]{2009A&A...493..409S}
{Simionescu}, A., {Werner}, N., {B{\"o}hringer}, H., {et~al.} 2009, \aap, 493,
  409

\bibitem[{{Smartt}(2009)}]{2009ARA&A..47...63S}
{Smartt}, S.~J. 2009, \araa, 47, 63

\bibitem[{{Sullivan} {et~al.}(2006){Sullivan}, {Le Borgne}, {Pritchet},
  {Hodsman}, {Neill}, {Howell}, {Carlberg}, {Astier}, {Aubourg}, {Balam},
  {Basa}, {Conley}, {Fabbro}, {Fouchez}, {Guy}, {Hook}, {Pain},
  {Palanque-Delabrouille}, {Perrett}, {Regnault}, {Rich}, {Taillet}, {Baumont},
  {Bronder}, {Ellis}, {Filiol}, {Lusset}, {Perlmutter}, {Ripoche}, \&
  {Tao}}]{2006ApJ...648..868S}
{Sullivan}, M., {Le Borgne}, D., {Pritchet}, C.~J., {et~al.} 2006, \apj, 648,
  868

\bibitem[{{Tamura} {et~al.}(2009){Tamura}, {Maeda}, {Mitsuda}, {Fabian},
  {Sanders}, {Furuzawa}, {Hughes}, {Iizuka}, {Matsushita}, \&
  {Tamagawa}}]{2009ApJ...705L..62T}
{Tamura}, T., {Maeda}, Y., {Mitsuda}, K., {et~al.} 2009, \apjl, 705, L62

\bibitem[{{Timmes} {et~al.}(1995){Timmes}, {Woosley}, \&
  {Weaver}}]{1995ApJS...98..617T}
{Timmes}, F.~X., {Woosley}, S.~E., \& {Weaver}, T.~A. 1995, \apjs, 98, 617

\bibitem[{{Tinsley}(1980)}]{1980FCPh....5..287T}
{Tinsley}, B.~M. 1980, \fcp, 5, 287

\bibitem[{{Treu} {et~al.}(2010){Treu}, {Auger}, {Koopmans}, {Gavazzi},
  {Marshall}, \& {Bolton}}]{2010ApJ...709.1195T}
{Treu}, T., {Auger}, M.~W., {Koopmans}, L.~V.~E., {et~al.} 2010, \apj, 709,
  1195

\bibitem[{{Tsujimoto} {et~al.}(1995){Tsujimoto}, {Nomoto}, {Yoshii},
  {Hashimoto}, {Yanagida}, \& {Thielemann}}]{1995MNRAS.277..945T}
{Tsujimoto}, T., {Nomoto}, K., {Yoshii}, Y., {et~al.} 1995, \mnras, 277, 945

\bibitem[{{Umeda} \& {Nomoto}(2002)}]{2002ApJ...565..385U}
{Umeda}, H. \& {Nomoto}, K. 2002, \apj, 565, 385

\bibitem[{{Waldman} {et~al.}(2011){Waldman}, {Sauer}, {Livne}, {Perets},
  {Glasner}, {Mazzali}, {Truran}, \& {Gal-Yam}}]{2011ApJ...738...21W}
{Waldman}, R., {Sauer}, D., {Livne}, E., {et~al.} 2011, \apj, 738, 21

\bibitem[{{Webbink}(1984)}]{1984ApJ...277..355W}
{Webbink}, R.~F. 1984, \apj, 277, 355

\bibitem[{{Werner} {et~al.}(2006){Werner}, {de Plaa}, {Kaastra}, {Vink},
  {Bleeker}, {Tamura}, {Peterson}, \& {Verbunt}}]{2006A&A...449..475W}
{Werner}, N., {de Plaa}, J., {Kaastra}, J.~S., {et~al.} 2006, \aap, 449, 475

\bibitem[{{Werner} {et~al.}(2008){Werner}, {Durret}, {Ohashi}, {Schindler}, \&
  {Wiersma}}]{2008SSRv..134..337W}
{Werner}, N., {Durret}, F., {Ohashi}, T., {Schindler}, S., \& {Wiersma},
  R.~P.~C. 2008, \ssr, 134, 337

\bibitem[{{Whelan} \& {Iben}(1973)}]{1973ApJ...186.1007W}
{Whelan}, J. \& {Iben}, Jr., I. 1973, \apj, 186, 1007

\bibitem[{{Woosley} \& {Weaver}(1995)}]{1995ApJS..101..181W}
{Woosley}, S.~E. \& {Weaver}, T.~A. 1995, \apjs, 101, 181

\bibitem[{{Yamaguchi} {et~al.}(2015){Yamaguchi}, {Badenes}, {Foster}, {Bravo},
  {Williams}, {Maeda}, {Nobukawa}, {Eriksen}, {Brickhouse}, {Petre}, \&
  {Koyama}}]{2015ApJ...801L..31Y}
{Yamaguchi}, H., {Badenes}, C., {Foster}, A.~R., {et~al.} 2015, \apjl, 801, L31

\bibitem[{{Yasumi} {et~al.}(2014){Yasumi}, {Nobukawa}, {Nakashima}, {Uchida},
  {Sugawara}, {Tsuru}, {Tanaka}, \& {Koyama}}]{2014PASJ...66...68Y}
{Yasumi}, M., {Nobukawa}, M., {Nakashima}, S., {et~al.} 2014, \pasj, 66, 68

\bibitem[{{Yates} {et~al.}(2016){Yates}, {Thomas}, \&
  {Henriques}}]{2016arXiv160304858Y}
{Yates}, R.~M., {Thomas}, P.~A., \& {Henriques}, B.~M.~B. 2016, ArXiv e-prints

\bibitem[{{Yoshii} {et~al.}(1996){Yoshii}, {Tsujimoto}, \&
  {Nomoto}}]{1996ApJ...462..266Y}
{Yoshii}, Y., {Tsujimoto}, T., \& {Nomoto}, K. 1996, \apj, 462, 266

\bibitem[{{Yuan} {et~al.}(2013){Yuan}, {Kobayashi}, {Schmidt}, {Podsiadlowski},
  {Sim}, \& {Scalzo}}]{2013MNRAS.432.1680Y}
{Yuan}, F., {Kobayashi}, C., {Schmidt}, B.~P., {et~al.} 2013, \mnras, 432, 1680

\end{thebibliography}
\bibliographystyle{aa}




\newpage

\appendix

\section{The effect of electron capture rates on the SNIa nucleosynthesis yields}\label{sect:ecapture}

Because the electron gas in the core of WDs is highly degenerate, the electron capture process can play an important role during the SNIa explosion, and has a significant impact on the nucleosynthesis of intermediate-mass and Fe-group elements. The Classical SNIa yield models referred to in this paper are taken from \citet{1999ApJS..125..439I}, who used the electron capture rates of \citet{1982ApJS...48..279F}. These rates were tabulated for light (i.e. $sd$-shell) nuclei only. Later on, \citet{2000ApJ...536..934B} showed that updated calculations of heavier ($pf$-shell) nuclei \citep[e.g.][]{1998PhLB..436...19L,2001ADNDT..79....1L} lead to a significant reduction in the electron capture rates compared to the previous estimates. In principle, these lowered rates affect the overall nucleosynthesis predicted by the SNIa models.

\begin{figure}[!]
        \centering
                \includegraphics[width=0.49\textwidth]{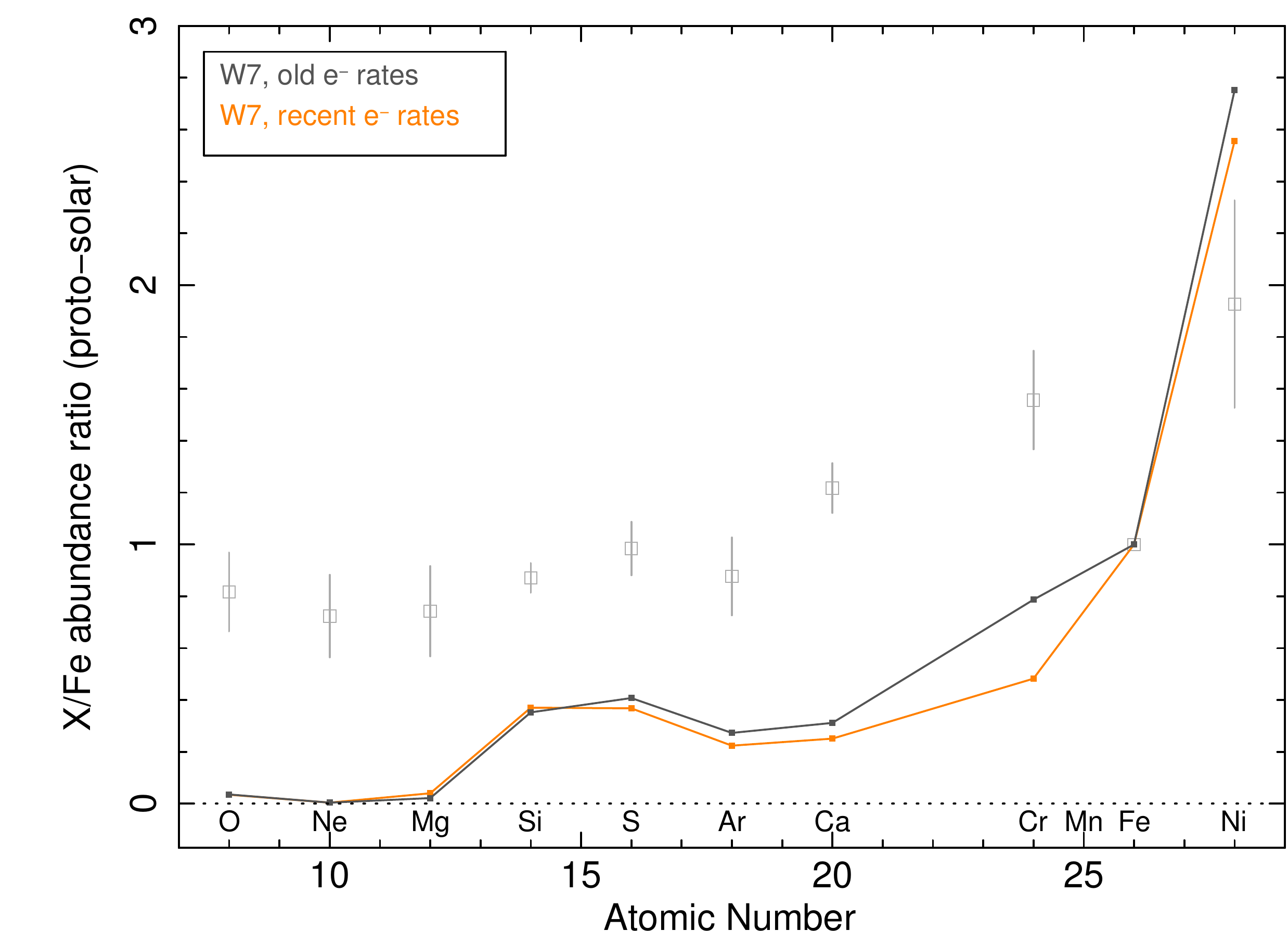}

        \caption{Predicted X/Fe abundances from the Classical W7 model (SNIa), adopted from \citet[][brown]{1999ApJS..125..439I} and \citet[][orange]{2010ApJ...712..624M}. The two models assume different electron capture rates, leading to different X/Fe ratios, both for intermediate-mass and Fe-group elements (see text). For comparison, the ICM average abundance ratios (inferred from Paper I) are also plotted.}
\label{fig:ecapture}
\end{figure}

In Fig. \ref{fig:ecapture}, we compare the X/Fe ratios predicted by the Classical W7 model, using first the older and then the more recent electron capture rates. These two W7 models are directly adopted from \citet{1999ApJS..125..439I} and \citet{2010ApJ...712..624M}, respectively. The largest difference in the X/Fe ratios from the more recent calculations is found for Cr/Fe, with a decrease of 39\% compared to the older electron capture rates. The other abundance ratios, however, show less pronounced differences ($\sim$20\% at most).

Except W7, no other Classical (or Bravo) model incorporating these updated electron capture rates is explicitly available in the literature. Although we do not expect this issue to alter the conclusions of this paper, this illustrates well that SN yield models may suffer from uncertainties, and that care must be taken when interpreting the ability of the models to strictly reproduce the measured abundance ratios in the ICM.

\onecolumn

\section{List of SN yield models used in this work}

\longtab{
\begin{longtable}{lccl}
\caption{\label{table:SNe_models} SNIa and SNcc yield models, taken from literature and used in this work.} \\
   \hline
   \hline
   Category & Name & Reference & Remarks \\
  \hline
   \endfirsthead
\caption{Continued.}\\
   \hline
   \hline
   Category & Name & Reference & Remarks \\
  \hline
  \endhead
  \hline
   \endfoot 
\multicolumn{4}{c}{SNIa}  \\    
\hline
Classical & W7 & 1 & Deflagration, $\rho_9 = 2.12$ \\
Classical & W70 & 1 & Deflagration, $\rho_9 = 2.12$, zero initial metallicity \\
Classical & WDD1 & 1 & Delayed-detonation, $\rho_9 = 2.12$, $\rho_\text{T,7} = 1.7$ \\
Classical & WDD2 & 1 & Delayed-detonation, $\rho_9 = 2.12$, $\rho_\text{T,7} = 2.2$ \\
Classical & WDD3 & 1 & Delayed-detonation, $\rho_9 = 2.12$, $\rho_\text{T,7} = 3.0$ \\
Classical & CDD1 & 1 & Delayed-detonation, $\rho_9 = 1.37$, $\rho_\text{T,7} = 1.7$ \\
Classical & CDD2 & 1 & Delayed-detonation, $\rho_9 = 1.37$, $\rho_\text{T,7} = 2.2$ \\
Bravo & DDTa & 2,3 & Delayed-detonation, fits the Tycho SNR, $\rho_\text{T,7} = 3.9$ \\
Bravo & DDTc & 2,3 & Delayed-detonation, fits the Tycho SNR, $\rho_\text{T,7} = 2.2$ \\
Bravo & DDTe & 2,3 & Delayed-detonation, fits the Tycho SNR, $\rho_\text{T,7} = 1.3$ \\
Ca-rich gap & CO.45HE.2 & 4 & Ca-rich SNe, $M_\text{CO} = 0.45$, $M_\text{He} = 0.2$ \\
Ca-rich gap & CO.5HE.2 & 4 & Ca-rich SNe, $M_\text{CO} = 0.5$, $M_\text{He} = 0.2$ \\
Ca-rich gap & CO.5HE.15 & 4 & Ca-rich SNe, $M_\text{CO} = 0.5$, $M_\text{He} = 0.15$ \\
Ca-rich gap & CO.5HE.2N.02 & 4 & Ca-rich SNe, $M_\text{CO} = 0.5$, $M_\text{He} = 0.2$, 2\% N in He layer \\
Ca-rich gap & CO.5HE.2C.03 & 4 & Ca-rich SNe, $M_\text{CO} = 0.5$, $M_\text{He} = 0.2$, 30\% mixing core-He layer \\
Ca-rich gap & CO.5HE.3 & 4 & Ca-rich SNe, $M_\text{CO} = 0.5$, $M_\text{He} = 0.3$ \\
Ca-rich gap & CO.55HE.2 & 4 & Ca-rich SNe, $M_\text{CO} = 0.55$, $M_\text{He} = 0.2$ \\
Ca-rich gap & CO.6HE.2 & 4 & Ca-rich SNe, $M_\text{CO} = 0.6$, $M_\text{He} = 0.2$ \\
2D & C-DEF & 5 & 2D deflagration, $\rho_9 = 2.9$ \\
2D & C-DDT & 5 & 2D delayed-detonation, $\rho_9 = 2.9$, $\rho_\text{T,7} = 1.0$ \\
2D & O-DDT & 5 & 2D delayed-detonation, $\rho_9 = 2.9$, $\rho_\text{T,7} = 1.0$, off-centre ignition \\
3D & N1def & 6 & 3D deflagration, $\rho_9 = 2.9$, 1 ignition spot \\
3D & N3def & 6 & 3D deflagration, $\rho_9 = 2.9$, 3 ignition spots \\
3D & N5def & 6 & 3D deflagration, $\rho_9 = 2.9$, 5 ignition spots \\
3D & N10def & 6 & 3D deflagration, $\rho_9 = 2.9$, 10 ignition spots \\
3D & N20def & 6 & 3D deflagration, $\rho_9 = 2.9$, 20 ignition spots \\
3D & N40def & 6 & 3D deflagration, $\rho_9 = 2.9$, 40 ignition spots \\
3D & N100Ldef & 6 & 3D deflagration, $\rho_9 = 1.0$, 100 ignition spots \\
3D & N100def & 6 & 3D deflagration, $\rho_9 = 2.9$, 100 ignition spots \\
3D & N100Hdef & 6 & 3D deflagration, $\rho_9 = 5.5$, 100 ignition spots \\
3D & N150def & 6 & 3D deflagration, $\rho_9 = 2.9$, 150 ignition spots \\
3D & N200def & 6 & 3D deflagration, $\rho_9 = 2.9$, 200 ignition spots \\
3D & N300Cdef & 6 & 3D deflagration, $\rho_9 = 2.9$, 300 centred ignition spots \\
3D & N1600def & 6 & 3D deflagration, $\rho_9 = 2.9$, 1600 ignition spots \\
3D & N1600Cdef & 6 & 3D deflagration, $\rho_9 = 2.9$, 1600 centred ignition spots \\
3D & N1 & 7 &  3D delayed-detonation, $\rho_9 = 2.9$, 1 ignition spot \\
3D & N3 & 7 &  3D delayed-detonation, $\rho_9 = 2.9$, 3 ignition spots \\
3D & N5 & 7 &  3D delayed-detonation, $\rho_9 = 2.9$, 5 ignition spots \\
3D & N10 & 7 &  3D delayed-detonation, $\rho_9 = 2.9$, 10 ignition spots \\
3D & N20 & 7 &  3D delayed-detonation, $\rho_9 = 2.9$, 20 ignition spots \\
3D & N40 & 7 &  3D delayed-detonation, $\rho_9 = 2.9$, 40 ignition spots \\
3D & N100L & 7 &  3D delayed-detonation, $\rho_9 = 1.0$, 100 ignition spots \\
3D & N100 & 7 &  3D delayed-detonation, $\rho_9 = 2.9$, 100 ignition spots \\
3D & N100H & 7 &  3D delayed-detonation, $\rho_9 = 5.5$, 100 ignition spots \\
3D & N150 & 7 &  3D delayed-detonation, $\rho_9 = 2.9$, 150 ignition spots \\
3D & N200 & 7 &  3D delayed-detonation, $\rho_9 = 2.9$, 200 ignition spots \\
3D & N300C & 7 &  3D delayed-detonation, $\rho_9 = 2.9$, 300 centred ignition spots \\
3D & N1600 & 7 &  3D delayed-detonation, $\rho_9 = 2.9$, 1600 ignition spots \\
3D & N1600C & 7 &  3D delayed-detonation, $\rho_9 = 2.9$, 1600 centred ignition spots \\
Sub-$M_\text{Ch}$ & 0.9\_0.9 & 8 & WD-WD violent merger, $M_\text{WD} \simeq 0.9$, $\rho_9 = 1.4\times 10^{-2}$ \\
\hline
\multicolumn{4}{c}{SNcc}  \\    
\hline
Nomoto & Z0 & 9,10,11 & $Z_\text{init} = 0$ \\
Nomoto & Z0\_cut & 9,10,11 & $Z_\text{init} = 0$, restricted to $\le$40 $M_\sun$ \\
Nomoto & Z0.001 & 9,10,11 & $Z_\text{init} = 0.001$ \\
Nomoto & Z0.004 & 9,10,11 & $Z_\text{init} = 0.004$ \\
Nomoto & Z0.008 & 11 & $Z_\text{init} = 0.008$ \\
Nomoto & Z0.02 & 9,10,11 & $Z_\text{init} = 0.02$ \\
Nomoto & Z0+PISNe & 9,10,11,12 & $Z_\text{init} = 0$, incl. contribution from PISNe (up to 300 $M_\sun$) \\
HW & Z0+PISNe & 13,14 & $Z_\text{init} = 0$, incl. contribution from PISNe (up to 260 $M_\sun$) \\

\end{longtable}
\tablefoot{The inner core densities $\rho_9$ are given in units of $10^9$ g/cm$^3$. The transitional deflagration-to-detonation densities $\rho_{T,7}$ are given in units of $10^7$ g/cm$^3$. The masses of the CO core and of the He layer (respectively $M_\text{CO}$ and $M_\text{He}$, "Ca-rich gap" models), and the mass of each of the two merging WD ($M_\text{WD}$, "DD channel" model) are given in units of $M_\sun$.}\\

\tablebib{(1)~\citet{1999ApJS..125..439I};
(2) \citet{2003ApJ...593..358B}; (3) \citet{2006ApJ...645.1373B}; (4) \citet{2011ApJ...738...21W};
(5) \citet{2010ApJ...712..624M}; (6) \citet{2014MNRAS.438.1762F}; (7) \citet{2013MNRAS.429.1156S};
(8) \citet{2010Natur.463...61P}; (9) \citet{2006NuPhA.777..424N}; (10) \citet{2006ApJ...653.1145K};
(11) \citet{2013ARA&A..51..457N}; (12) \citet{2002ApJ...565..385U}; (13) \citet{2002ApJ...567..532H};
(14) \citet{2010ApJ...724..341H}.
}
}

\end{document}